\documentclass[aps,prl,longbibliography,reprint,showpacs,superscriptaddress,amsmath]{revtex4-1}
\usepackage{color}
\usepackage{graphicx}
\begin{document}

\title{Effect of biaxial strain on the phase transitions of Ca(Fe$_{1-x}$Co$_x$)$_2$As$_2$}

\author{A. E. B\"ohmer}  
\affiliation{Ames Laboratory US DOE, Ames, Iowa 50011, USA}

\author{A. Sapkota}  
\affiliation{Ames Laboratory US DOE, Ames, Iowa 50011, USA}
\affiliation{Department of Physics and Astronomy, Iowa State University, Ames, Iowa 50011, USA}
\author{A. Kreyssig}  
\affiliation{Ames Laboratory US DOE, Ames, Iowa 50011, USA}
\affiliation{Department of Physics and Astronomy, Iowa State University, Ames, Iowa 50011, USA}
\author{S. L. Bud'ko}  
\affiliation{Ames Laboratory US DOE, Ames, Iowa 50011, USA}
\affiliation{Department of Physics and Astronomy, Iowa State University, Ames, Iowa 50011, USA}
\author{G. Drachuck}  
\affiliation{Ames Laboratory US DOE, Ames, Iowa 50011, USA}
\affiliation{Department of Physics and Astronomy, Iowa State University, Ames, Iowa 50011, USA}
\author{S. M. Saunders}  
\affiliation{Ames Laboratory US DOE, Ames, Iowa 50011, USA}
\affiliation{Department of Physics and Astronomy, Iowa State University, Ames, Iowa 50011, USA}
\author{A. I. Goldman} 
\affiliation{Ames Laboratory US DOE, Ames, Iowa 50011, USA}
\affiliation{Department of Physics and Astronomy, Iowa State University, Ames, Iowa 50011, USA} 
\author{P. C. Canfield}
\affiliation{Ames Laboratory US DOE, Ames, Iowa 50011, USA}
\affiliation{Department of Physics and Astronomy, Iowa State University, Ames, Iowa 50011, USA}

\date{\today}

\begin{abstract}

 We study the effect of applied strain as a physical control parameter for the phase transitions of Ca(Fe$_{1-x}$Co$_x$)$_2$As$_2$ using resistivity, magnetization, x-ray diffraction and $^{57}$Fe M\"ossbauer spectroscopy. Biaxial strain, namely compression of the basal plane of the tetragonal unit cell, is created through firm bonding of samples to a rigid substrate, via differential thermal expansion. This strain is shown to induce a magneto-structural phase transition in originally paramagnetic samples; and superconductivity in previously non-superconducting ones. The magneto-structural transition is gradual as a consequence of using strain instead of pressure or stress as a tuning parameter.

\end{abstract}


\maketitle

Tuning parameters are an essential tool in the study of correlated materials, since they can selectively promote specific interactions. As an example, unconventional superconductivity often emerges around the point where antiferromagnetic order is suppressed by hydrostatic pressure \cite{Uemura2009}. Strain has been occasionally used as a tuning parameter \cite{Stillwell1968,Overcash1969,Angadi1973,Gaidukov1977,Shayegan2003}, but is less widely employed than pressure. Recently, new piezo-based strain-tuning devices have been presented \cite{Hicks2014II,Gannon2015} and used in the study of ruthenates \cite{Hicks2014,Steppke2016} and SmB$_6$ \cite{Stern2016}. Additionally, strain has been famously employed to probe the nematic susceptibility of iron-based superconductors \cite{Chu2012,Kuo2013,Kuo2016,Shapiro2015,He2016}. 
Applying strain means enforcing a deformation, or length change with respect to a 'free' reference state, and can be achieved using a rigid device. Notably, strain directly affects the electronic band structure and properties. Such enforced deformations depend on fewer elastic constants than deformations achieved by applying force (stress, pressure).

The iron-based superconductors \cite{Canfield2010,Johnston2010,Si2016} sport a complex and highly tunable interplay between antiferromagnetism (AFM), a tetragonal-to-orthorhombic structural distortion and superconductivity. The variety of suitable tuning parameters includes diverse chemical substitutions \cite{Canfield2010,Merz2016}, hydrostatic pressure \cite{Canfield2010,Takahashi2008,Sefat2011}, epitaxial strain in thin films \cite{Iida2009,Engelmann2013,Kazumasa2016} and uniaxial pressure \cite{Torikachvili2009,Prokes2010,Budko2009,Yamazaki2010,Meingast2012}.
In Ca(Fe$_{1-x}$Co$_x$)$_2$As$_2$, substitution of Co for Fe suppresses a coupled first-order magneto-structural transition at $T_{\mathrm{s,N}}$ and induces superconductivity with a maximum $T_c$ of 16 K \cite{Ran2012}. Ca(Fe$_{1-x}$Co$_x$)$_2$As$_2$ is exceptionally pressure-sensitive  \cite{Torikachvili2009,Prokes2010,Gati2012,Budko2012}, as 
 exemplified by the unprecedentedly large rate of suppression of $T_{\mathrm{s,N}}$ with hydrostatic pressure, $\mathrm{d}T_{\mathrm{s,N}}/\mathrm{d}p\approx-1100$ K/GPa ($x=0.028$) \cite{Gati2012}, two orders of magnitude larger than for BaFe$_2$As$_2$ \cite{Colombier2009}, and by the sensitivity of the material to post-growth treatment \cite{Ran2011,Ran2012,Ranthesis}.

Here, we study the effect of strain on Ca(Fe$_{1-x}$Co$_x$)$_2$As$_2$ with a combination of macroscopic and microscopic probes. Biaxial in-plane strain is achieved by making use of the differential thermal expansion between the samples and a rigid substrate, to which the samples are firmly bonded. 
It directly affects the $c/a$ ratio of the tetragonal samples, similarly to uniaxial pressure along the $c$ direction. In contrast to uniaxial pressure along the tetragonal [110]  direction, commonly used for detwinning in iron-based systems \cite{Fisher2011}, it does not break the tetragonal symmetry. We demonstrate that the $c/a$ ratio is a suitable tuning parameter for the phase transitions of Ca(Fe$_{1-x}$Co$_x$)$_2$As$_2$.

\begin{figure}
	\includegraphics[width=8.6cm]{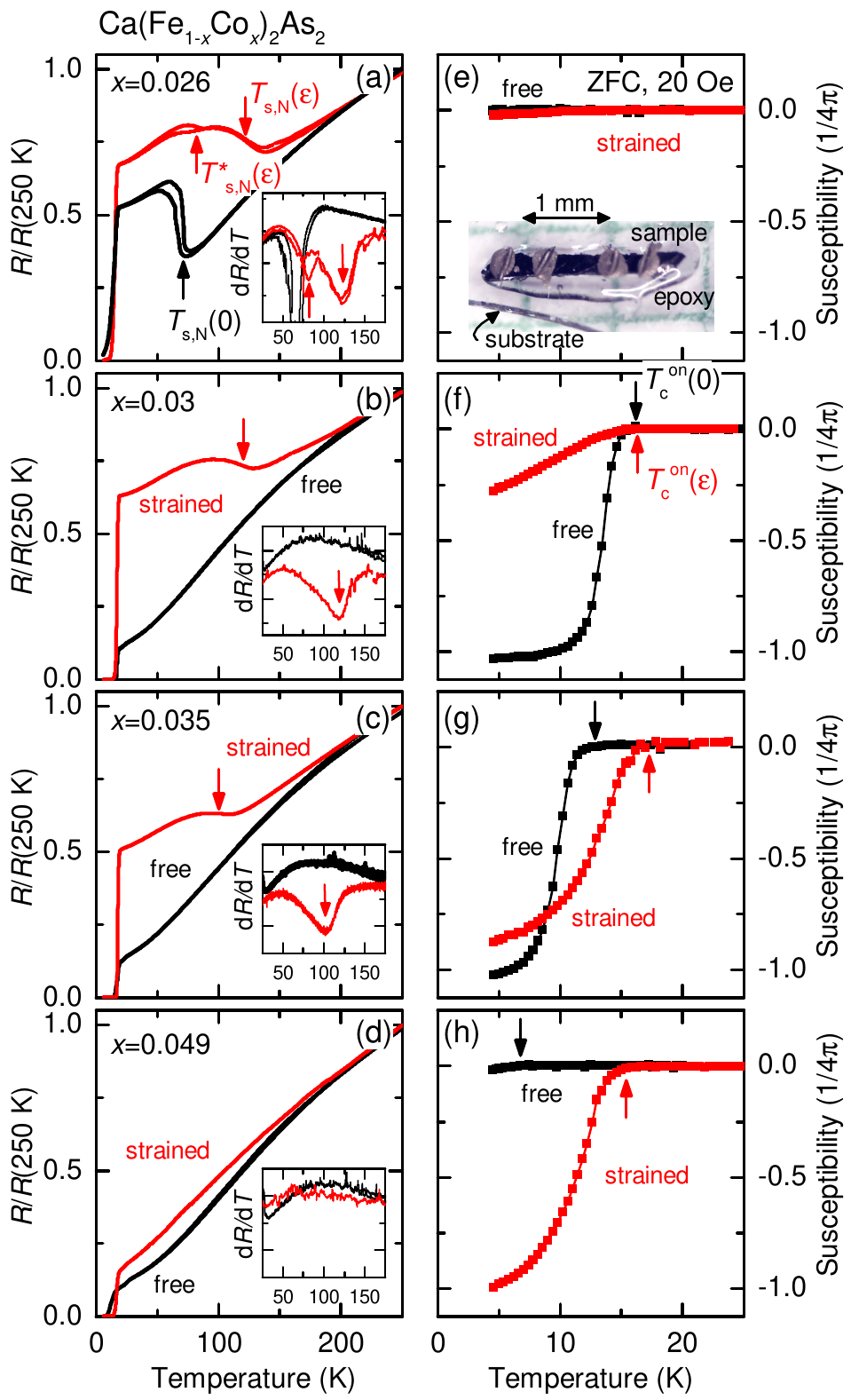}
	\caption{(a)-(d) Normalized electrical resistivity and (e)-(h) zero-field cooled magnetization (superconducting shielding) of Ca(Fe$_{1-x}$Co$_x$)$_2$As$_2$, with varying Co content $x$. Measurements were performed on the same thin bar-shaped sample--first free (`$0$') and then strained (`$\varepsilon$') by gluing to the glass substrate [photograph in the inset in (e)]. Magnetization was measured parallel sample length to minimize demagnetization effects.  }
	\label{fig:1}
\end{figure}

Samples of Ca(Fe$_{1-x}$Co$_x$)$_2$As$_2$ ($0\leq x\leq 0.054$, with $x$ determined by wave-length dispersive x-ray spectroscopy) were grown out of FeAs flux and annealed at 400$^\circ$C, ensuring the absence of the collapsed-tetragonal phase present in as-grown samples \cite{Ran2011,Ran2012,Ranthesis}. 
 Samples for resistivity and magnetization measurements were cleaved and cut into small thin bars of typical dimensions of $\sim 1.5\times0.2\times0.05$ mm$^3$ and mass of $0.2-0.5$ mg. To create strain ($\varepsilon$), samples were glued with Devcon 5 minute epoxy to a piece of thin borosilicate glass (Fisherbrand Cover Glass, 160 $\mu$m thickness) as shown in the inset to Fig. 1(e). Electrical resistance was measured with an LS370 AC resistance bridge. Magnetic susceptibility was measured under zero-field cooled (ZFC) conditions in a Quantum Design MPMS SQUID magnetometer. $^{57}$Fe M\"ossbauer spectroscopy was performed in transmission using a SEE Co. conventional constant acceleration type spectrometer with a $^{57}$Co(Rh) source kept at room temperature on a set of $\sim40$ samples [$x=0.035$, typical dimensions of $2\times1\times(0.04-0.1)$ mm$^3$]. High-energy (100.3 keV) x-ray diffraction was performed similarly to in Ref. \onlinecite{Sapkota2014}, on a strained sample from the M\"ossbauer set, employing a Pixirad-1 detector. Thermal expansion was also measured with a home-built capacitance dilatometer \cite{Schmiedeshoff2006}.

Figure \ref{fig:1} shows the normalized resistivity and the ZFC magnetization of a selection of samples with different Co concentrations. Each sample was first measured in free-standing and then in strained conditions to directly reveal the impact of strain. The data on free samples are in very good agreement with previous work \cite{Ran2012,Ranthesis}, yet strain induces dramatic changes. 
For the underdoped, $x=0.026$, sample, the sharp rise of resistivity at $T_{\mathrm{s,N}}(0)\approx70$ K in the free state, is replaced by a broader anomaly at a higher temperature $T_{\mathrm{s,N}}(\varepsilon)\approx125$ K [defined as the minimum in derivative, insets in Figs. 1(a)-(c)] and a hysteretic anomaly at $T^*_{\mathrm{s,N}}(\varepsilon)\approx80$ K in the strained state. No significant superconducting shielding fraction is observed in either state, consistent with the mutual exclusion of superconductivity and antiferromagnetism in Ca(Fe$_{1-x}$Co$_x$)$_2$As$_2$ \cite{Ran2012}. Free-standing samples having slightly higher Co content, $x=0.03$ and $x=0.035$, show no magneto-structural transition and have full superconducting shielding with onset $T_c^{\mathrm{on}}$ values of 15.5 K and 12 K, respectively. Under strain, clear anomalies in the resistance appear at $T_{\mathrm{s,N}}(\varepsilon)=120$ K and 100 K, respectively, and the superconducting shielding fraction decreases. For the highest Co content ($x=0.049$), in contrast, the strain induces no high-temperature anomaly. Instead, strain induces full superconducting shielding in this sample that showed only a tiny trace of superconductivity in its free state.
	
In order to characterize microscopically the strain and the strain-induced resistivity anomalies, we performed high-energy x-ray diffraction on strained Ca(Fe$_{0.965}$Co$_{0.035}$)$_2$As$_2$, and compare the results with the uniaxial thermal expansion of free samples as determined by capacitance dilatometry (Fig. 2). 
The in-plane length of free-standing Ca(Fe$_{0.965}$Co$_{0.035}$)$_2$As$_2$ [solid grey line in Fig. 2(a)] increases strongly upon decreasing temperature, whereas the thermal expansion of the glass substrate is fairly low (solid blue line). The diffraction data shows that the in-plane axis of strained Ca(Fe$_{0.965}$Co$_{0.035}$)$_2$As$_2$ follows the substrate length rather closely, which means that it is compressed with respect to the free standing state for $T\gtrsim100$ K. As seen from Fig. 2 (b), the strained sample's $c$-axis is expanded with respect to its length in the free-standing state, as is expected from the Poisson effect \cite{Tomic2012}. Thus, the imposed strain corresponds to an elongation of the tetragonal unit cell, by $\left[L_{a,b}(\varepsilon)-L_{a,b}(0)\right]/L_{a,b}(0)\sim-0.3$\% and $\left[L_{c}(\varepsilon)-L_{c}(0)\right]/L_{c}(0)\sim0.55$\% at 105 K, as shown schematically in the right inset of Fig. 2(a). 

Since the high-energy x-rays penetrate the entire sample thickness, the strain distribution may be inferred from the diffraction data (colormaps in Fig. 2). The narrow intensity distribution around $a=5.51$ $\AA$ at 105 K implies homogeneous strain in the $\sim1\times10^{-3}$ mm$^3$ sample volume illuminated by the $0.1\times0.1$ mm$^2$ x-ray beam, oriented perpendicular to the sample surface. In the $c$-axis measurement [(0 0 10) reflection] the intensity is peaked around $c=11.52$ $\AA$ at 105 K but has a tail towards lower values, indicating that a fraction of the $\sim1\times10^{-2}$ mm$^3$  sample volume illuminated by the beam at the small angle of $\sim5^\circ$ experiences reduced strain.

 \begin{figure}
 	\includegraphics[width=8.6cm]{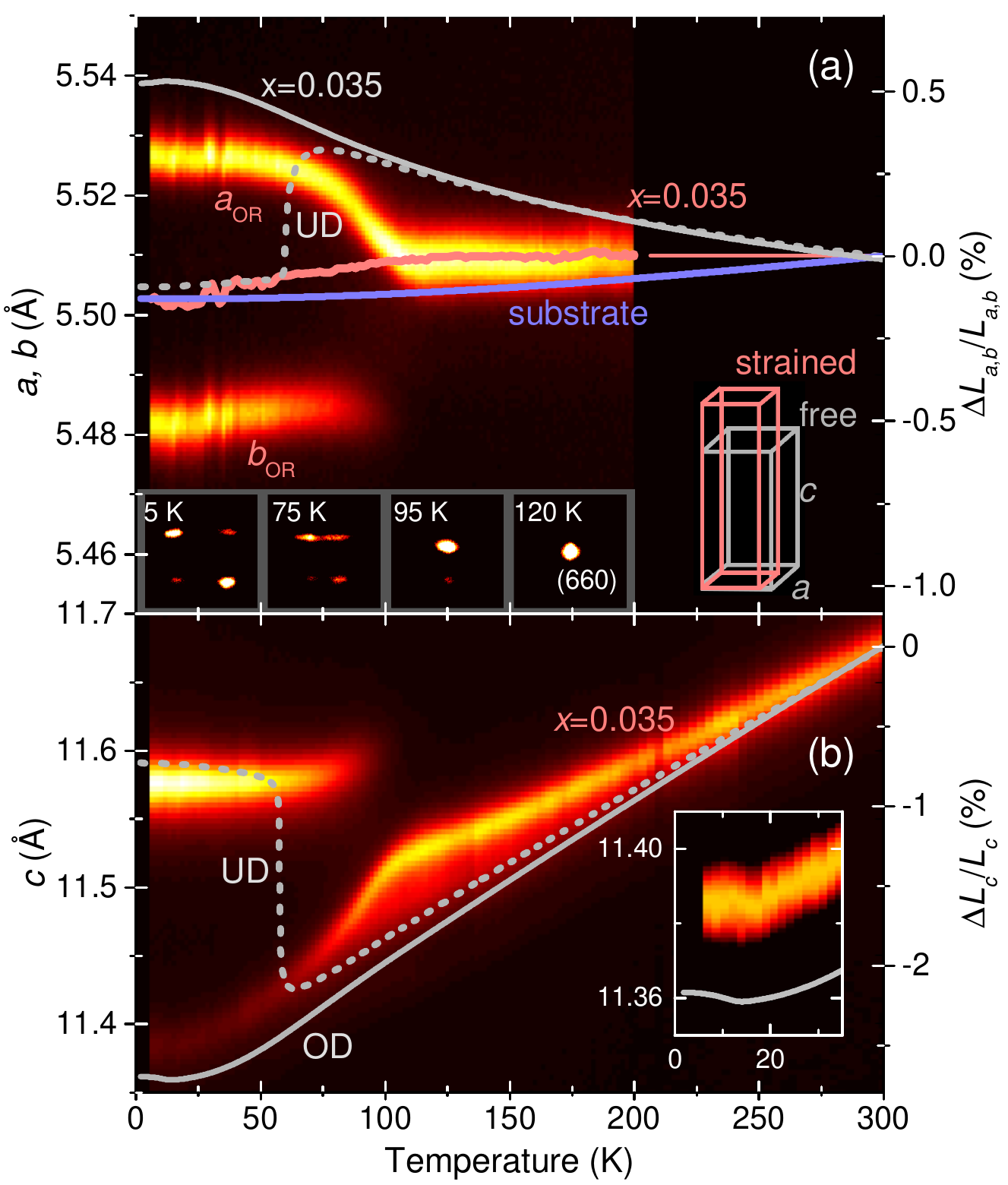}
 	\caption{(a) in-plane and (b) $c$-axis structural data for Ca(Fe$_{1-x}$Co$_x$)$_2$As$_2$. Lattice parameters of strained Ca(Fe$_{0.965}$Co$_{0.035}$)$_2$As$_2$ from x-ray diffraction measured on warming are shown as color-coded intensity maps (left axis). Lines indicate uniaxial fractional length changes, $\Delta L_i/L_i$ ($i=c$, $c$ axis and  $i=a,b$, in-plane average), of free overdoped (OD) samples [(a) $x=0.035$ (this work) and (b) $x=0.029$ \cite{Budko2012}] and of a representative underdoped (UD) $x=0.027$ sample \cite{Budko2012} obtained by capacitance dilatometry [right axis, scaled so that $\Delta L_c/L_c$ corresponds to the lattice parameter change $[c(T)-c(300\textnormal{ K})]/c(300 \textnormal{ K})$ in (b) and analogously in (a)]. The blue line in (a) shows the substrate thermal expansion and the red line indicates the average in-plane length of strained Ca(Fe$_{0.965}$Co$_{0.035}$)$_2$As$_2$ inferred from the diffraction data. The right inset in (a) depicts schematically the deformation of the unit cell in the strained state. The row of insets in (a) shows the $(HK0)$ diffraction pattern close to the tetragonal $(660)$ reflection revealing orthorhombic domains. The inset in (b) presents the data on expanded scales. 
 }
 	\label{fig:2}
 \end{figure}

 \begin{figure}
 	\includegraphics[width=8.6cm]{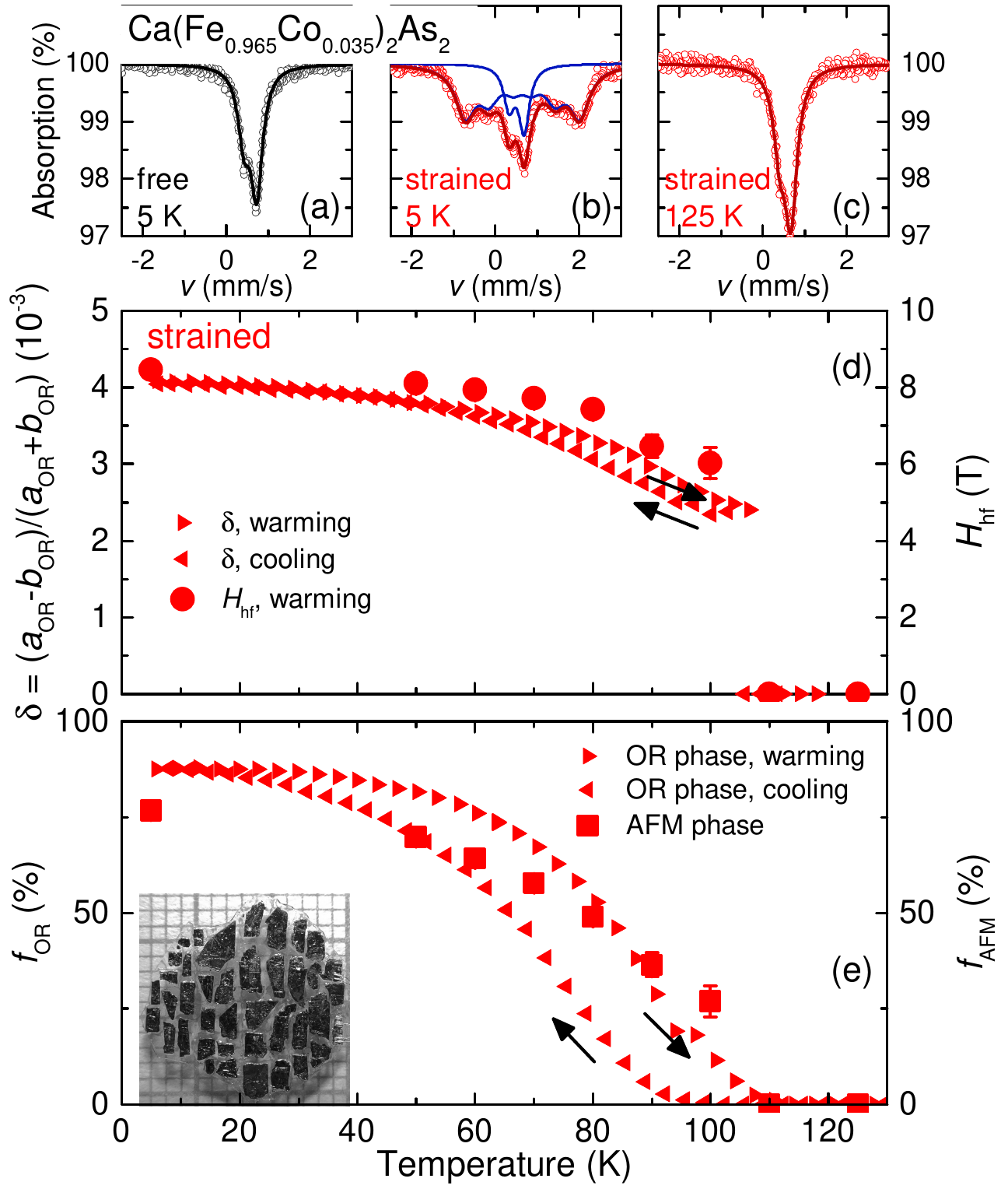}
 	\caption{(a)-(c) Examples of $^{57}$Fe M\"ossbauer spectra of Ca(Fe$_{0.965}$Co$_{0.035}$)$_2$As$_2$ (symbols) and the fitted paramagnetic doublet and magnetic sextet (lines). (d) Orthorhombic order parameter $\delta=\left(a_{\mathrm{OR}}-b_{\mathrm{OR}}\right)/\left(a_{\mathrm{OR}}+b_{\mathrm{OR}}\right)$ and magnetic hyperfine field $H_{\mathrm{hf}}$. (e) AFM and OR phase fractions, $f_{\mathrm{AFM}}$ and $f_{\mathrm{OR}}$, respectively. $f_{\mathrm{AFM}}$, deduced from M\"ossbauer spectroscopy (which probes the whole sample mosaic) reaches a maximum of $\sim80\%$ and $f_{\mathrm{OR}}$ deduced from $c$-axis x-ray diffraction (which probes only a part of the same) reaches $\sim90\%$. Diffraction reveals a temperature hysteresis, whereas M\"ossbauer spectroscopy was performed only on warming. The inset shows a photograph of the samples used for these measurements on a mm grid.}
 	\label{fig:2b}
 \end{figure}

 The tetragonal-to-orthorhombic (T-to-OR) structural transition is obvious from the split of the in-plane lattice parameter into $a_{\mathrm {OR}}$ and $b_{\mathrm {OR}}$ in Fig. 2(a) and is, in more detail, shown by the splitting of the tetragonal (6 6 0) reflection, similar to what is seen in $A$Fe$_2$As$_2$ ($A$=Ba, Sr, Ca) parent compounds \cite{Tanatar2009}. Note that a globally firm bonding between sample and substrate at all temperatures is supported by the observation that the center of the diffraction pattern, i.e., the average in-plane length of the sample as inferred from x-ray diffraction, follows the substrate length quite closely.
 
A peculiarity is that in the strained $x=0.035$ sample, two phase fractions coexist from 105 K down to the lowest temperature, as clearly visible in the $c$-axis diffraction data. The `transformed' OR phase fraction, $f_{\mathrm{OR}}$, has a significantly larger $c$-lattice parameter, comparable to the increase of the $c$ axis of free underdoped samples on entering the OR/AFM phase. The `remaining' T phase fraction has a smaller $c$-lattice parameter, which appears to exhibit a small kink at $\sim15$ K [visualized in the inset to Fig. 2(b)], reminiscent of the signature of bulk superconductivity in free overdoped samples \cite{Budko2012}.

  \begin{figure}
  	\includegraphics[width=8.6cm]{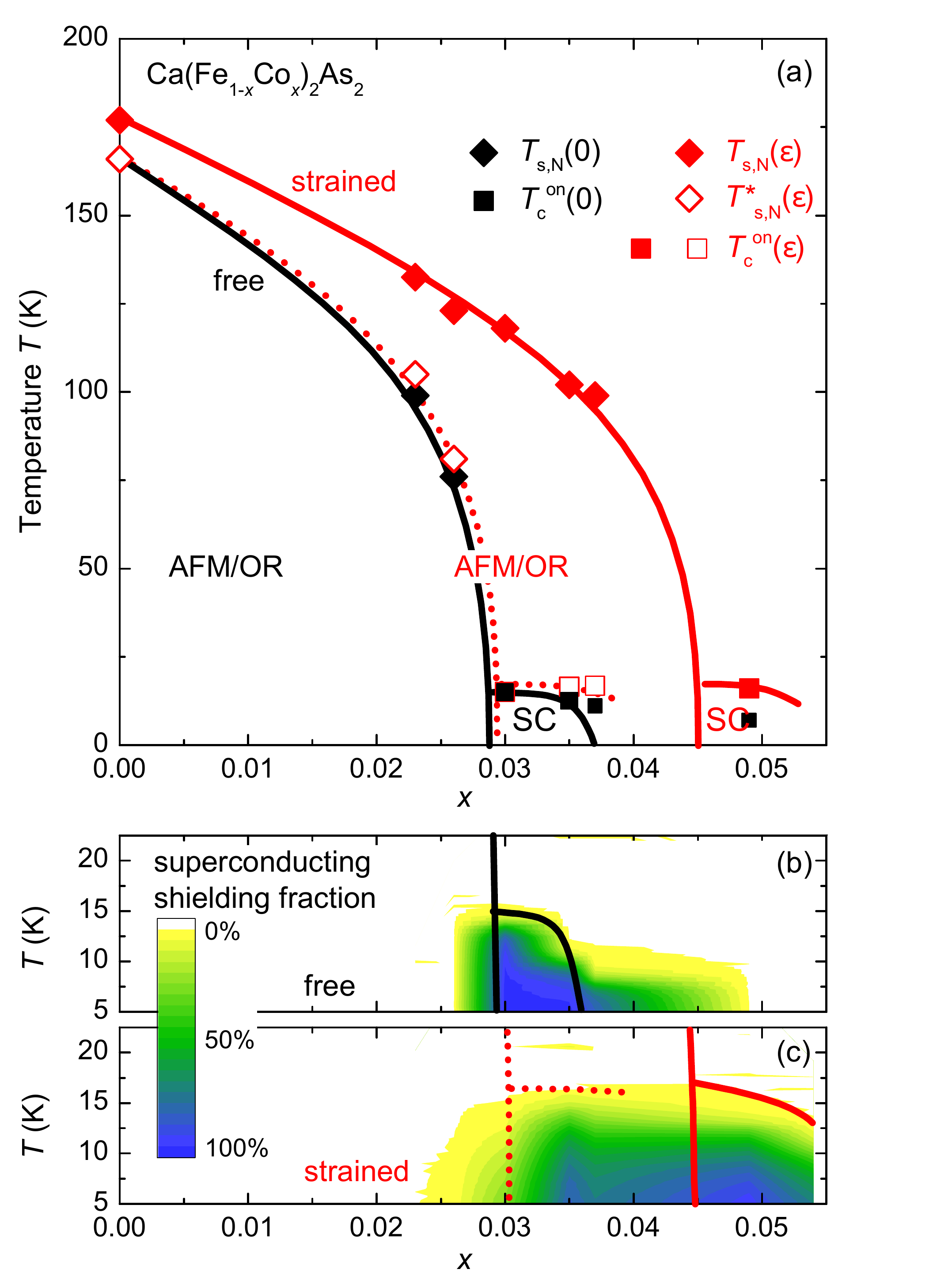}
  	\caption{(a) Phase diagram of Ca(Fe$_{1-x}$Co$_x$)$_2$As$_2$ in the free (black) and strained (red) state with increased $c/a$ ratio. The AFM/OR transition at $T_{\mathrm{s,N}}(\varepsilon)$ is only gradual. Red open symbols and dashed lines correspond to the remaining phase fraction within the strained sample. (b),(c) Superconducting shielding fraction of free and strained samples, respectively. Lines are a guide to the eye.}
  	\label{fig:4}
  	
  \end{figure}

  The following simplified picture likely explains the peculiarities of the strain-induced first-order transition in our samples. 
   Firm bonding to the rigid substrate forces the in-plane sample length $L_{\mathrm{sample}}$ to be equal to a length $L_{\mathrm{substrate}}$. On decreasing temperature, the sample experiences increasing strain: the basal plane is compressed, and the $c$ axis expands relative to the free state. This kind of deformation favors the OR phase thermodynamically, since its average in-plane length, $L_{\mathrm{OR}}$, is smaller than in the T phase ($L_{\mathrm{T}}$), and its $c$-axis length is larger (dashed lines in Fig. 2) \cite{Budko2012}. At a critical temperature/value of strain, the OR phase nucleates in some parts of the sample. Those transformed parts have a reduced in-plane area so that some of the strain is released. Hence, the remaining T phase can partly relax its lattice parameters and does not transform. The boundary condition is expressed by   $L_{\mathrm{sample}}=f_{\mathrm{OR}}L_{\mathrm{OR}}+\left(1-f_{\mathrm{OR}}\right)L_{\mathrm{T}}=L_{\mathrm{substrate}}$
   and, if $L_{\mathrm{OR}}<L_{\mathrm{substrate}}<L_{\mathrm{T}}$, a solution that balances the free energies of the T and the OR phase, and the elastic energy of the lattice deformations, likely entails $0<f_{\mathrm{OR}}<1$. As a consequence, there is a well-defined phase coexistence and $f_{\textnormal{OR}}$ changes gradually with temperature. Note that such a phase coexistence implies locally inhomogeneous strain. 
   
   The relaxation of the lattice parameters of the remaining T phase below the onset of the transition is clearly visible as kinks in the diffraction data around $\sim$100 K. 
   The $a$ axis of the remaining T phase follows rather closely the orthorhombic $a_{\mathrm{OR}}$ axis, and is therefore not distinctly visible at low temperatures. 
     Finally, for underdoped samples, the yet untransformed phase fraction will naturally undergo the AFM/OR transition on its own close to the transition temperature under free-standing conditions, which explains the second anomaly, at $T^*_{\mathrm{s,N}}(\varepsilon)\approx T_{\mathrm{s,N}}(0)$, in resistivity in Fig. 1(a). 

Whether the induced OR phase in strained Ca(Fe$_{0.965}$Co$_{0.035}$)$_2$As$_2$ is also magnetically ordered, is studied using $^{57}$Fe M\"ossbauer spectroscopy, a local probe of the ordered magnetic hyperfine field (Fig. \ref{fig:2b}). 
 The doublet-type spectrum measured on free samples (attached to the glass substrate with Apiezon N-grease) confirms their paramagnetic ground state. In contrast, when the same samples are glued with epoxy to the glass substrate and thus strained, the spectrum is given by a superposition of a paramagnetic doublet and a magnetic sextet. The relative areas indicate that a fraction of $f_{\textnormal{AFM}}\approx80\%$ of the Fe nuclei experience a distinct magnetic hyperfine field at low temperatures. On increasing temperature, a purely paramagnetic state is recovered at 125 K. 
Figures \ref{fig:2b}(d),(e) summarize the obtained structural distortion and magnetic hyperfine field, as well as the respective ordered phase fractions. The low-temperature values of both the magnetic hyperfine field $H_{\mathrm{hf}}$ and orthorhombic distortion $\delta$ are only $\sim20\%$ lower than the values of pure CaFe$_2$As$_2$, namely $H_{\mathrm hf}=10$ T \cite{Ran2011} and $\delta(0)=5 \times 10^{-3}$ \cite{Goldman2008}. In addition, $\delta(T)$ and $H_{\mathrm{hf}}(T)$ follow each other closely, indicating that a coupled first-order magneto-structural transition is indeed induced by the strain. 
	
The phase diagram in Fig. 4 is constructed from the resistivity and magnetization data. The superconducting shielding fraction is presented as color-coded maps in panels (b) and (c). In agreement with previous reports \cite{Ran2012}, free-standing samples show a steep doping-induced suppression of the magneto-structural transition at $x\approx0.028$ and a superconducting half-dome between $x=0.03-0.035$. The strained samples exhibit a significantly extended range of the AFM/OR phase, as confirmed by the microscopic probes. The superconducting shielding is reduced in this range and full shielding is reached only for $x=0.049$. Hence, the superconducting dome is less sharply defined than for free samples. This is a natural consequence of the phase coexistence under strain. At low temperatures, the remaining tetragonal phase fraction becomes superconducting, while the OR phase fraction likely stays non-superconducting. Note that the ZFC shielding fraction may overestimate the true superconducting volume fraction.

In summary, increasing the $c/a$ ratio through applying biaxial strain, shifts the phase diagram of Ca(Fe$_{1-x}$Co$_x$)$_2$As$_2$ to higher $x$, without a change of the maximum $T_c$. This suggests the possibility that the maximum $T_c$ in this system has already been reached. 
In general, the initial rate of change of transition temperatures with uniaxial pressure can be inferred from thermodynamic relations, using, in particular, uniaxial thermal expansion. The trends inferred for Ca(Fe$_{1-x}$Co$_x$)$_2$As$_2$  \cite{Gati2012, Budko2012} agree with the present result, whereas a quantitative estimate that takes into account the compressibility of the material agrees within a factor of $\sim 2-3$. 
Notably, in most iron-based systems, uniaxial pressure derivatives have opposite sign along the $a$ and $c$ axes \cite{Budko2012,Meingast2012,Boehmer2015II}, indicating that these systems are notably more sensitive to changes of the $c/a$ ratio than to hydrostatic pressure, which averages those components and entails partial cancellation of opposing effects.
As studied in detail in BaFe$_2$As$_2$, however, the relation between the phase diagram and changes of the $c/a$ ratio, when achieved either by pressure or by substitution with various transition metals, is non-universal \cite{Ni2009, Thaler2010}.

This strain-tuning is analogous to epitaxial strain in thin films, which was recently studied in thin films of Ba(Fe$_{1-x}$Co$_x$)$_2$As$_2$ grown on different substrates. An in-plane strain of almost $\pm0.6\%$ was achieved and yielded changes of $T_\textnormal{N}$ and $T_\textnormal{c}$ of $\lesssim10$ K \cite{Kazumasa2016}. The changes are very similar to the present result, considering a lower strain-sensitivity of BaFe$_2$As$_2$ with respect to CaFe$_2$As$_2$.
 The magnitude of applied strain in this study depends on temperature and on the difference in thermal expansivity between sample and substrate. We expect that
 BaFe$_2$As$_2$ rigidly glued to our glass substrate experiences only $\sim-0.02$\% of  in-plane strain at $T_\textnormal{s,N}=140$ K, but $\sim-0.12$\% when a copper substrate is used. This should result in a small but measurable shift of $T_\textnormal{s,N}$ by $1-2$ K. The effect on $T_c$ of optimally doped Ba(Fe$_{1-x}$Co$_x$)$_2$As$_2$ is likely of similar size. 

At any strain-induced first-order phase transition, however, the change of lattice parameters will result in (partial) strain release and a well-defined phase coexistence, similarly to our observations. In this respect, controlling the total sample length is fundamentally different from controlling stress or pressure. 
This has implications for other techniques utilizing strain and needs to be taken into account when gluing thin samples to a rigid substrate, which is a common practice in a variety of experimental techniques.

In conclusion, biaxial strain is established as a tuning parameter for the phase transitions of bulk Ca(Fe$_{1-x}$Co$_x$)$_2$As$_2$. Although the observed strain effects are particularly pronounced in the extremely sensitive CaFe$_2$As$_2$ system, they can occur in principle in any material.

We are grateful to D. S. Robinson for the excellent support of
the x-ray diffraction measurements and to Valentin Taufour, Makariy Tanatar and Herman Suderow for helpful discussions. 
This work was supported by the
U.S. Department
of Energy (DOE), Office of Basic Energy Science,
Division of Materials Sciences and Engineering. Ames Laboratory is
operated for the DOE by Iowa State
University under Contract No. DE-AC02-07CH11358. A. E. B. acknowledges support from the Helmholtz Association via PD-226. G. D. was supported by
the Gordon and Betty Moore Foundations EPiQS Initiative
through Grant GBMF4411. This research used
resources of the Advanced Photon Source, a DOE Office of Science User Facility operated for
the DOE Office of Science by Argonne National Laboratory
under Contract No. DE-AC02-06CH11357.

%


\begin{thebibliography}{45}%
	\makeatletter
	\providecommand \@ifxundefined [1]{%
		\@ifx{#1\undefined}
	}%
	\providecommand \@ifnum [1]{%
		\ifnum #1\expandafter \@firstoftwo
		\else \expandafter \@secondoftwo
		\fi
	}%
	\providecommand \@ifx [1]{%
		\ifx #1\expandafter \@firstoftwo
		\else \expandafter \@secondoftwo
		\fi
	}%
	\providecommand \natexlab [1]{#1}%
	\providecommand \enquote  [1]{``#1''}%
	\providecommand \bibnamefont  [1]{#1}%
	\providecommand \bibfnamefont [1]{#1}%
	\providecommand \citenamefont [1]{#1}%
	\providecommand \href@noop [0]{\@secondoftwo}%
	\providecommand \href [0]{\begingroup \@sanitize@url \@href}%
	\providecommand \@href[1]{\@@startlink{#1}\@@href}%
	\providecommand \@@href[1]{\endgroup#1\@@endlink}%
	\providecommand \@sanitize@url [0]{\catcode `\\12\catcode `\$12\catcode
		`\&12\catcode `\#12\catcode `\^12\catcode `\_12\catcode `\%12\relax}%
	\providecommand \@@startlink[1]{}%
	\providecommand \@@endlink[0]{}%
	\providecommand \url  [0]{\begingroup\@sanitize@url \@url }%
	\providecommand \@url [1]{\endgroup\@href {#1}{\urlprefix }}%
	\providecommand \urlprefix  [0]{URL }%
	\providecommand \Eprint [0]{\href }%
	\providecommand \doibase [0]{http://dx.doi.org/}%
	\providecommand \selectlanguage [0]{\@gobble}%
	\providecommand \bibinfo  [0]{\@secondoftwo}%
	\providecommand \bibfield  [0]{\@secondoftwo}%
	\providecommand \translation [1]{[#1]}%
	\providecommand \BibitemOpen [0]{}%
	\providecommand \bibitemStop [0]{}%
	\providecommand \bibitemNoStop [0]{.\EOS\space}%
	\providecommand \EOS [0]{\spacefactor3000\relax}%
	\providecommand \BibitemShut  [1]{\csname bibitem#1\endcsname}%
	\let\auto@bib@innerbib\@empty
	\bibitem [{\citenamefont {Uemura}(2009)}]{Uemura2009}%
	\BibitemOpen
	\bibfield  {author} {\bibinfo {author} {\bibfnamefont {Yasutomo~J.}\
			\bibnamefont {Uemura}},\ }\bibfield  {title} {\enquote {\bibinfo {title}
			{Superconductivity: {C}ommonalities in phase and mode},}\ }\href {\doibase
		10.1038/nmat2415} {\bibfield  {journal} {\bibinfo  {journal} {Nature
				Materials}\ }\textbf {\bibinfo {volume} {8}},\ \bibinfo {pages} {253--255}
		(\bibinfo {year} {2009})}\BibitemShut {NoStop}%
	\bibitem [{\citenamefont {E.~P.~Stillwell}\ and\ \citenamefont
		{Davis}(1968)}]{Stillwell1968}%
	\BibitemOpen
	\bibfield  {author} {\bibinfo {author} {\bibfnamefont {M.~J.~Skove}\
			\bibnamefont {E.~P.~Stillwell}}\ and\ \bibinfo {author} {\bibfnamefont
			{J.~H.}\ \bibnamefont {Davis}},\ }\bibfield  {title} {\enquote {\bibinfo
			{title} {Two {"whisker"} straining devices suitable for low temperatures},}\
	}\href@noop {} {\bibfield  {journal} {\bibinfo  {journal} {Rev. Sci.
				Instrum.}\ }\textbf {\bibinfo {volume} {39}},\ \bibinfo {pages} {155}
		(\bibinfo {year} {1968})}\BibitemShut {NoStop}%
	\bibitem [{\citenamefont {Overcash}\ \emph {et~al.}(1969)\citenamefont
		{Overcash}, \citenamefont {Skove},\ and\ \citenamefont
		{Stillwell}}]{Overcash1969}%
	\BibitemOpen
	\bibfield  {author} {\bibinfo {author} {\bibfnamefont {D.~R.}\ \bibnamefont
			{Overcash}}, \bibinfo {author} {\bibfnamefont {M.~J.}\ \bibnamefont {Skove}},
		\ and\ \bibinfo {author} {\bibfnamefont {E.~P.}\ \bibnamefont {Stillwell}},\
	}\bibfield  {title} {\enquote {\bibinfo {title} {Effect of elastic stress on
				some electronic properties of indium},}\ }\href@noop {} {\bibfield  {journal}
		{\bibinfo  {journal} {Physical Review}\ }\textbf {\bibinfo {volume} {187}},\
		\bibinfo {pages} {570--574} (\bibinfo {year} {1969})}\BibitemShut {NoStop}%
	\bibitem [{\citenamefont {Angadi}\ \emph {et~al.}(1973)\citenamefont {Angadi},
		\citenamefont {Britton},\ and\ \citenamefont {Fawcett}}]{Angadi1973}%
	\BibitemOpen
	\bibfield  {author} {\bibinfo {author} {\bibfnamefont {M.~A.}\ \bibnamefont
			{Angadi}}, \bibinfo {author} {\bibfnamefont {D.~E.}\ \bibnamefont {Britton}},
		\ and\ \bibinfo {author} {\bibfnamefont {E.}~\bibnamefont {Fawcett}},\
	}\bibfield  {title} {\enquote {\bibinfo {title} {Low temperature sample
				holder for rotating a crystal under tension in a superconducting soleniod},}\
	}\href@noop {} {\bibfield  {journal} {\bibinfo  {journal} {Journal of Physics
				E: Scientific instruments}\ }\textbf {\bibinfo {volume} {6}},\ \bibinfo
		{pages} {1086--1087} (\bibinfo {year} {1973})}\BibitemShut {NoStop}%
	\bibitem [{\citenamefont {Ga\u{i}dukov}\ \emph {et~al.}(1977)\citenamefont
		{Ga\u{i}dukov}, \citenamefont {Danilova},\ and\ \citenamefont
		{Shcherbina-Saimo\u{i}lova}}]{Gaidukov1977}%
	\BibitemOpen
	\bibfield  {author} {\bibinfo {author} {\bibfnamefont {Yu.~P.}\ \bibnamefont
			{Ga\u{i}dukov}}, \bibinfo {author} {\bibfnamefont {N.~P.}\ \bibnamefont
			{Danilova}}, \ and\ \bibinfo {author} {\bibfnamefont {M.~B.}\ \bibnamefont
			{Shcherbina-Saimo\u{i}lova}},\ }\bibfield  {title} {\enquote {\bibinfo
			{title} {Phase transition of order 2 1/2 in zinc},}\ }\href@noop {}
	{\bibfield  {journal} {\bibinfo  {journal} {Pis'ma Zh. Eksp. Teor. Fiz.}\
		}\textbf {\bibinfo {volume} {25}},\ \bibinfo {pages} {509--513} (\bibinfo
		{year} {1977})}\BibitemShut {NoStop}%
	\bibitem [{\citenamefont {Shayegan}\ \emph {et~al.}(2003)\citenamefont
		{Shayegan}, \citenamefont {Karrai}, \citenamefont {Shkolnikov}, \citenamefont
		{Vakili}, \citenamefont {Poortere},\ and\ \citenamefont
		{Manus}}]{Shayegan2003}%
	\BibitemOpen
	\bibfield  {author} {\bibinfo {author} {\bibfnamefont {M.}~\bibnamefont
			{Shayegan}}, \bibinfo {author} {\bibfnamefont {K.}~\bibnamefont {Karrai}},
		\bibinfo {author} {\bibfnamefont {Y.~P.}\ \bibnamefont {Shkolnikov}},
		\bibinfo {author} {\bibfnamefont {K.}~\bibnamefont {Vakili}}, \bibinfo
		{author} {\bibfnamefont {E.~P.~De}\ \bibnamefont {Poortere}}, \ and\ \bibinfo
		{author} {\bibfnamefont {S.}~\bibnamefont {Manus}},\ }\bibfield  {title}
	{\enquote {\bibinfo {title} {Low-temperature, in situ tunable, uniaxial
				stress measurements in semiconductors using a piezoelectric actuator},}\
	}\href {\doibase 10.1063/1.1635963} {\bibfield  {journal} {\bibinfo
			{journal} {Appl. Phys. Lett.}\ }\textbf {\bibinfo {volume} {83}},\ \bibinfo
		{pages} {5235} (\bibinfo {year} {2003})}\BibitemShut {NoStop}%
	\bibitem [{\citenamefont {Hicks}\ \emph
		{et~al.}(2014{\natexlab{a}})\citenamefont {Hicks}, \citenamefont {Barber},
		\citenamefont {Edkins}, \citenamefont {Brodsky},\ and\ \citenamefont
		{Mackenzie}}]{Hicks2014II}%
	\BibitemOpen
	\bibfield  {author} {\bibinfo {author} {\bibfnamefont {C.W.}\ \bibnamefont
			{Hicks}}, \bibinfo {author} {\bibfnamefont {M.E.}\ \bibnamefont {Barber}},
		\bibinfo {author} {\bibfnamefont {S.D.}\ \bibnamefont {Edkins}}, \bibinfo
		{author} {\bibfnamefont {D.O.}\ \bibnamefont {Brodsky}}, \ and\ \bibinfo
		{author} {\bibfnamefont {A.P.}\ \bibnamefont {Mackenzie}},\ }\bibfield
	{title} {\enquote {\bibinfo {title} {Piezoelectric-based apparatus for strain
				tuning},}\ }\href {\doibase 10.1063/1.4881611} {\bibfield  {journal}
		{\bibinfo  {journal} {Review of Scientific Instruments}\ }\textbf {\bibinfo
			{volume} {85}},\ \bibinfo {pages} {065003} (\bibinfo {year}
		{2014}{\natexlab{a}})}\BibitemShut {NoStop}%
	\bibitem [{\citenamefont {Gannon}\ \emph {et~al.}(2015)\citenamefont {Gannon},
		\citenamefont {Bosak}, \citenamefont {Burkovsky}, \citenamefont {Nisbet},
		\citenamefont {Petrović},\ and\ \citenamefont {Hoesch}}]{Gannon2015}%
	\BibitemOpen
	\bibfield  {author} {\bibinfo {author} {\bibfnamefont {L.}~\bibnamefont
			{Gannon}}, \bibinfo {author} {\bibfnamefont {A.}~\bibnamefont {Bosak}},
		\bibinfo {author} {\bibfnamefont {R.~G.}\ \bibnamefont {Burkovsky}}, \bibinfo
		{author} {\bibfnamefont {G.}~\bibnamefont {Nisbet}}, \bibinfo {author}
		{\bibfnamefont {A.~P.}\ \bibnamefont {Petrović}}, \ and\ \bibinfo {author}
		{\bibfnamefont {M.}~\bibnamefont {Hoesch}},\ }\bibfield  {title} {\enquote
		{\bibinfo {title} {A device for the application of uniaxial strain to single
				crystal samples for use in synchrotron radiation experiments},}\ }\href
	{\doibase http://dx.doi.org/10.1063/1.4933383} {\bibfield  {journal}
		{\bibinfo  {journal} {Review of Scientific Instruments}\ }\textbf {\bibinfo
			{volume} {86}},\ \bibinfo {pages} {103904} (\bibinfo {year}
		{2015})}\BibitemShut {NoStop}%
	\bibitem [{\citenamefont {Hicks}\ \emph
		{et~al.}(2014{\natexlab{b}})\citenamefont {Hicks}, \citenamefont {Brodsky},
		\citenamefont {Yelland}, \citenamefont {Gibbs}, \citenamefont {Bruin},
		\citenamefont {Barber}, \citenamefont {Edkins}, \citenamefont {Nishimura},
		\citenamefont {Yonezawa}, \citenamefont {Maeno},\ and\ \citenamefont
		{Mackenzie}}]{Hicks2014}%
	\BibitemOpen
	\bibfield  {author} {\bibinfo {author} {\bibfnamefont {Clifford~W.}\
			\bibnamefont {Hicks}}, \bibinfo {author} {\bibfnamefont {Daniel~O.}\
			\bibnamefont {Brodsky}}, \bibinfo {author} {\bibfnamefont {Edward~A.}\
			\bibnamefont {Yelland}}, \bibinfo {author} {\bibfnamefont {Alexandra~S.}\
			\bibnamefont {Gibbs}}, \bibinfo {author} {\bibfnamefont {Jan A.~N.}\
			\bibnamefont {Bruin}}, \bibinfo {author} {\bibfnamefont {Mark~E.}\
			\bibnamefont {Barber}}, \bibinfo {author} {\bibfnamefont {Stephen~D.}\
			\bibnamefont {Edkins}}, \bibinfo {author} {\bibfnamefont {Keigo}\
			\bibnamefont {Nishimura}}, \bibinfo {author} {\bibfnamefont {Shingo}\
			\bibnamefont {Yonezawa}}, \bibinfo {author} {\bibfnamefont {Yoshiteru}\
			\bibnamefont {Maeno}}, \ and\ \bibinfo {author} {\bibfnamefont {Andrew~P.}\
			\bibnamefont {Mackenzie}},\ }\bibfield  {title} {\enquote {\bibinfo {title}
			{Strong increase of {$T_c$} of {Sr$_2$RuO$_4$} under both tensile and
				compressive strain},}\ }\href {\doibase 10.1126/science.1248292} {\bibfield
		{journal} {\bibinfo  {journal} {Science}\ }\textbf {\bibinfo {volume}
			{344}},\ \bibinfo {pages} {283--285} (\bibinfo {year}
		{2014}{\natexlab{b}})}\BibitemShut {NoStop}%
	\bibitem [{\citenamefont {{Steppke}}\ \emph {et~al.}(2016)\citenamefont
		{{Steppke}}, \citenamefont {{Zhao}}, \citenamefont {{Barber}}, \citenamefont
		{{Scaffidi}}, \citenamefont {{Jerzembeck}}, \citenamefont {{Rosner}},
		\citenamefont {{Gibbs}}, \citenamefont {{Maeno}}, \citenamefont {{Simon}},
		\citenamefont {{Mackenzie}},\ and\ \citenamefont {{Hicks}}}]{Steppke2016}%
	\BibitemOpen
	\bibfield  {author} {\bibinfo {author} {\bibfnamefont {A.}~\bibnamefont
			{{Steppke}}}, \bibinfo {author} {\bibfnamefont {L.}~\bibnamefont {{Zhao}}},
		\bibinfo {author} {\bibfnamefont {M.~E.}\ \bibnamefont {{Barber}}}, \bibinfo
		{author} {\bibfnamefont {T.}~\bibnamefont {{Scaffidi}}}, \bibinfo {author}
		{\bibfnamefont {F.}~\bibnamefont {{Jerzembeck}}}, \bibinfo {author}
		{\bibfnamefont {H.}~\bibnamefont {{Rosner}}}, \bibinfo {author}
		{\bibfnamefont {A.~S.}\ \bibnamefont {{Gibbs}}}, \bibinfo {author}
		{\bibfnamefont {Y.}~\bibnamefont {{Maeno}}}, \bibinfo {author} {\bibfnamefont
			{S.~H.}\ \bibnamefont {{Simon}}}, \bibinfo {author} {\bibfnamefont {A.~P.}\
			\bibnamefont {{Mackenzie}}}, \ and\ \bibinfo {author} {\bibfnamefont {C.~W.}\
			\bibnamefont {{Hicks}}},\ }\bibfield  {title} {\enquote {\bibinfo {title}
			{Strong peak in {T$_c$} of {Sr$_2$RuO$_4$} under uniaxial pressure},}\ }\href
	{http://adsabs.harvard.edu/abs/2016arXiv160406669S} {\bibfield  {journal}
		{\bibinfo  {journal} {ArXiv e-prints}\ } (\bibinfo {year} {2016})},\ \Eprint
	{http://arxiv.org/abs/1604.06669} {arXiv:1604.06669} \BibitemShut {NoStop}%
	\bibitem [{\citenamefont {{Stern}}\ \emph {et~al.}(2016)\citenamefont
		{{Stern}}, \citenamefont {{Dzero}}, \citenamefont {{Galitski}}, \citenamefont
		{{Fisk}},\ and\ \citenamefont {{Xia}}}]{Stern2016}%
	\BibitemOpen
	\bibfield  {author} {\bibinfo {author} {\bibfnamefont {A.}~\bibnamefont
			{{Stern}}}, \bibinfo {author} {\bibfnamefont {M.}~\bibnamefont {{Dzero}}},
		\bibinfo {author} {\bibfnamefont {V.~M.}\ \bibnamefont {{Galitski}}},
		\bibinfo {author} {\bibfnamefont {Z.}~\bibnamefont {{Fisk}}}, \ and\ \bibinfo
		{author} {\bibfnamefont {J.}~\bibnamefont {{Xia}}},\ }\bibfield  {title}
	{\enquote {\bibinfo {title} {Kondo insulator {SmB$_6$} under strain: surface
				dominated conduction near room temperature},}\ }\href@noop {} {\bibfield
		{journal} {\bibinfo  {journal} {ArXiv e-prints}\ } (\bibinfo {year}
		{2016})},\ \Eprint {http://arxiv.org/abs/1607.07454} {arXiv:1607.07454}
	\BibitemShut {NoStop}%
	\bibitem [{\citenamefont {Chu}\ \emph {et~al.}(2012)\citenamefont {Chu},
		\citenamefont {Kuo}, \citenamefont {Analytis},\ and\ \citenamefont
		{Fisher}}]{Chu2012}%
	\BibitemOpen
	\bibfield  {author} {\bibinfo {author} {\bibfnamefont {Jiun-Haw}\
			\bibnamefont {Chu}}, \bibinfo {author} {\bibfnamefont {Hsueh-Hui}\
			\bibnamefont {Kuo}}, \bibinfo {author} {\bibfnamefont {James~G.}\
			\bibnamefont {Analytis}}, \ and\ \bibinfo {author} {\bibfnamefont {Ian~R.}\
			\bibnamefont {Fisher}},\ }\bibfield  {title} {\enquote {\bibinfo {title}
			{Divergent nematic susceptibility in an iron arsenide superconductor},}\
	}\href {\doibase 10.1126/science.1221713} {\bibfield  {journal} {\bibinfo
			{journal} {Science}\ }\textbf {\bibinfo {volume} {337}},\ \bibinfo {pages}
		{710--712} (\bibinfo {year} {2012})}\BibitemShut {NoStop}%
	\bibitem [{\citenamefont {Kuo}\ \emph {et~al.}(2013)\citenamefont {Kuo},
		\citenamefont {Shapiro}, \citenamefont {Riggs},\ and\ \citenamefont
		{Fisher}}]{Kuo2013}%
	\BibitemOpen
	\bibfield  {author} {\bibinfo {author} {\bibfnamefont {Hsueh-Hui}\
			\bibnamefont {Kuo}}, \bibinfo {author} {\bibfnamefont {Maxwell~C.}\
			\bibnamefont {Shapiro}}, \bibinfo {author} {\bibfnamefont {Scott~C.}\
			\bibnamefont {Riggs}}, \ and\ \bibinfo {author} {\bibfnamefont {Ian~R.}\
			\bibnamefont {Fisher}},\ }\bibfield  {title} {\enquote {\bibinfo {title}
			{Measurement of the elastoresistivity coefficients of the underdoped iron
				arsenide {Ba(Fe$_{0.975}$Co$_{0.025}$)$_{2}$As$_{2}$}},}\ }\href {\doibase
		10.1103/PhysRevB.88.085113} {\bibfield  {journal} {\bibinfo  {journal} {Phys.
				Rev. B}\ }\textbf {\bibinfo {volume} {88}},\ \bibinfo {pages} {085113}
		(\bibinfo {year} {2013})}\BibitemShut {NoStop}%
	\bibitem [{\citenamefont {Kuo}\ \emph {et~al.}(2016)\citenamefont {Kuo},
		\citenamefont {Chu}, \citenamefont {Palmstrom}, \citenamefont {Kivelson},\
		and\ \citenamefont {Fisher}}]{Kuo2016}%
	\BibitemOpen
	\bibfield  {author} {\bibinfo {author} {\bibfnamefont {Hsueh-Hui}\
			\bibnamefont {Kuo}}, \bibinfo {author} {\bibfnamefont {Jiun-Haw}\
			\bibnamefont {Chu}}, \bibinfo {author} {\bibfnamefont {Johanna~C.}\
			\bibnamefont {Palmstrom}}, \bibinfo {author} {\bibfnamefont {Steven~A.}\
			\bibnamefont {Kivelson}}, \ and\ \bibinfo {author} {\bibfnamefont {Ian~R.}\
			\bibnamefont {Fisher}},\ }\bibfield  {title} {\enquote {\bibinfo {title}
			{Ubiquitous signatures of nematic quantum criticality in optimally doped
				{Fe}-based superconductors},}\ }\href {\doibase 10.1126/science.aab0103}
	{\bibfield  {journal} {\bibinfo  {journal} {Science}\ }\textbf {\bibinfo
			{volume} {352}},\ \bibinfo {pages} {958--962} (\bibinfo {year}
		{2016})}\BibitemShut {NoStop}%
	\bibitem [{\citenamefont {Shapiro}\ \emph {et~al.}(2015)\citenamefont
		{Shapiro}, \citenamefont {Hlobil}, \citenamefont {Hristov}, \citenamefont
		{Maharaj},\ and\ \citenamefont {Fisher}}]{Shapiro2015}%
	\BibitemOpen
	\bibfield  {author} {\bibinfo {author} {\bibfnamefont {M.~C.}\ \bibnamefont
			{Shapiro}}, \bibinfo {author} {\bibfnamefont {Patrik}\ \bibnamefont
			{Hlobil}}, \bibinfo {author} {\bibfnamefont {A.~T.}\ \bibnamefont {Hristov}},
		\bibinfo {author} {\bibfnamefont {Akash~V.}\ \bibnamefont {Maharaj}}, \ and\
		\bibinfo {author} {\bibfnamefont {I.~R.}\ \bibnamefont {Fisher}},\ }\bibfield
	{title} {\enquote {\bibinfo {title} {Symmetry constraints on the
				elastoresistivity tensor},}\ }\href {\doibase 10.1103/PhysRevB.92.235147}
	{\bibfield  {journal} {\bibinfo  {journal} {Phys. Rev. B}\ }\textbf {\bibinfo
			{volume} {92}},\ \bibinfo {pages} {235147} (\bibinfo {year}
		{2015})}\BibitemShut {NoStop}%
	\bibitem [{\citenamefont {{He}}\ \emph {et~al.}(2016)\citenamefont {{He}},
		\citenamefont {{Wang}}, \citenamefont {{Ahn}}, \citenamefont {{Hardy}},
		\citenamefont {{Wolf}}, \citenamefont {{Adelmann}}, \citenamefont
		{{Schmalian}}, \citenamefont {{Eremin}},\ and\ \citenamefont
		{{Meingast}}}]{He2016}%
	\BibitemOpen
	\bibfield  {author} {\bibinfo {author} {\bibfnamefont {M.}~\bibnamefont
			{{He}}}, \bibinfo {author} {\bibfnamefont {L.}~\bibnamefont {{Wang}}},
		\bibinfo {author} {\bibfnamefont {F.}~\bibnamefont {{Ahn}}}, \bibinfo
		{author} {\bibfnamefont {F.}~\bibnamefont {{Hardy}}}, \bibinfo {author}
		{\bibfnamefont {T.}~\bibnamefont {{Wolf}}}, \bibinfo {author} {\bibfnamefont
			{P.}~\bibnamefont {{Adelmann}}}, \bibinfo {author} {\bibfnamefont
			{J.}~\bibnamefont {{Schmalian}}}, \bibinfo {author} {\bibfnamefont
			{I.}~\bibnamefont {{Eremin}}}, \ and\ \bibinfo {author} {\bibfnamefont
			{C.}~\bibnamefont {{Meingast}}},\ }\bibfield  {title} {\enquote {\bibinfo
			{title} {Dichotomy between in-plane magnetic susceptibility and resistivity
				anisotropies in extremely strained {BaFe$_{2}$As$_{2}$}},}\ }\href
	{http://adsabs.harvard.edu/abs/2016arXiv161005575H} {\bibfield  {journal}
		{\bibinfo  {journal} {ArXiv e-prints}\ } (\bibinfo {year} {2016})},\ \Eprint
	{http://arxiv.org/abs/1610.05575} {1610.05575} \BibitemShut {NoStop}%
	\bibitem [{\citenamefont {Canfield}\ and\ \citenamefont
		{Bud'ko}(2010)}]{Canfield2010}%
	\BibitemOpen
	\bibfield  {author} {\bibinfo {author} {\bibfnamefont {Paul~C.}\ \bibnamefont
			{Canfield}}\ and\ \bibinfo {author} {\bibfnamefont {Sergey~L.}\ \bibnamefont
			{Bud'ko}},\ }\bibfield  {title} {\enquote {\bibinfo {title} {{FeAs}-based
				superconductivity: {A} case study of the effects of transition metal doping
				on {BaFe$_2$As$_2$}},}\ }\href {\doibase
		10.1146/annurev-conmatphys-070909-104041} {\bibfield  {journal} {\bibinfo
			{journal} {Annual Review of Condensed Matter Physics}\ }\textbf {\bibinfo
			{volume} {1}},\ \bibinfo {pages} {27--50} (\bibinfo {year}
		{2010})}\BibitemShut {NoStop}%
	\bibitem [{\citenamefont {Johnston}(2010)}]{Johnston2010}%
	\BibitemOpen
	\bibfield  {author} {\bibinfo {author} {\bibfnamefont {David~C.}\
			\bibnamefont {Johnston}},\ }\bibfield  {title} {\enquote {\bibinfo {title}
			{The puzzle of high temperature superconductivity in layered iron pnictides
				and chalcogenides},}\ }\href {\doibase 10.1080/00018732.2010.513480}
	{\bibfield  {journal} {\bibinfo  {journal} {Advances in Physics}\ }\textbf
		{\bibinfo {volume} {59}},\ \bibinfo {pages} {803--1061} (\bibinfo {year}
		{2010})}\BibitemShut {NoStop}%
	\bibitem [{\citenamefont {Si}\ \emph {et~al.}(2016)\citenamefont {Si},
		\citenamefont {Yu},\ and\ \citenamefont {Abrahams}}]{Si2016}%
	\BibitemOpen
	\bibfield  {author} {\bibinfo {author} {\bibfnamefont {Qimiao}\ \bibnamefont
			{Si}}, \bibinfo {author} {\bibfnamefont {Rong}\ \bibnamefont {Yu}}, \ and\
		\bibinfo {author} {\bibfnamefont {Elihu}\ \bibnamefont {Abrahams}},\
	}\bibfield  {title} {\enquote {\bibinfo {title} {High-temperature
				superconductivity in iron pnictides and chalcogenides},}\ }\href {\doibase
		10.1038/natrevmats.2016.17} {\bibfield  {journal} {\bibinfo  {journal}
			{Nature Reviews Materials}\ }\textbf {\bibinfo {volume} {1}},\ \bibinfo
		{pages} {16017} (\bibinfo {year} {2016})}\BibitemShut {NoStop}%
	\bibitem [{\citenamefont {Merz}\ \emph {et~al.}(2016)\citenamefont {Merz},
		\citenamefont {Schweiss}, \citenamefont {Nagel}, \citenamefont {Huang},
		\citenamefont {Eder}, \citenamefont {Wolf}, \citenamefont {von L\"ohneysen},\
		and\ \citenamefont {Schuppler}}]{Merz2016}%
	\BibitemOpen
	\bibfield  {author} {\bibinfo {author} {\bibfnamefont {Michael}\ \bibnamefont
			{Merz}}, \bibinfo {author} {\bibfnamefont {Peter}\ \bibnamefont {Schweiss}},
		\bibinfo {author} {\bibfnamefont {Peter}\ \bibnamefont {Nagel}}, \bibinfo
		{author} {\bibfnamefont {Meng-Jie}\ \bibnamefont {Huang}}, \bibinfo {author}
		{\bibfnamefont {Robert}\ \bibnamefont {Eder}}, \bibinfo {author}
		{\bibfnamefont {Thomas}\ \bibnamefont {Wolf}}, \bibinfo {author}
		{\bibfnamefont {Hilbert}\ \bibnamefont {von L\"ohneysen}}, \ and\ \bibinfo
		{author} {\bibfnamefont {Stefan}\ \bibnamefont {Schuppler}},\ }\bibfield
	{title} {\enquote {\bibinfo {title} {Of substitution and doping: Spatial and
				electronic structure in {Fe} pnictides},}\ }\href {\doibase
		10.7566/JPSJ.85.044707} {\bibfield  {journal} {\bibinfo  {journal} {Journal
				of the Physical Society of Japan}\ }\textbf {\bibinfo {volume} {85}},\
		\bibinfo {pages} {044707} (\bibinfo {year} {2016})}\BibitemShut {NoStop}%
	\bibitem [{\citenamefont {Takahashi}\ \emph {et~al.}(2008)\citenamefont
		{Takahashi}, \citenamefont {Igawa}, \citenamefont {Arii}, \citenamefont
		{Kamihara}, \citenamefont {Hirano},\ and\ \citenamefont
		{Hosono}}]{Takahashi2008}%
	\BibitemOpen
	\bibfield  {author} {\bibinfo {author} {\bibfnamefont {Hiroki}\ \bibnamefont
			{Takahashi}}, \bibinfo {author} {\bibfnamefont {Kazumi}\ \bibnamefont
			{Igawa}}, \bibinfo {author} {\bibfnamefont {Kazunobu}\ \bibnamefont {Arii}},
		\bibinfo {author} {\bibfnamefont {Yoichi}\ \bibnamefont {Kamihara}}, \bibinfo
		{author} {\bibfnamefont {Masahiro}\ \bibnamefont {Hirano}}, \ and\ \bibinfo
		{author} {\bibfnamefont {Hideo}\ \bibnamefont {Hosono}},\ }\bibfield  {title}
	{\enquote {\bibinfo {title} {Superconductivity at {43 K} in an iron-based
				layered compound {LaO$_{1-x}$F$_x$FeAs}},}\ }\href {\doibase
		10.1038/nature06972} {\bibfield  {journal} {\bibinfo  {journal} {Nature}\
		}\textbf {\bibinfo {volume} {453}},\ \bibinfo {pages} {376--378} (\bibinfo
		{year} {2008})}\BibitemShut {NoStop}%
	\bibitem [{\citenamefont {Sefat}(2011)}]{Sefat2011}%
	\BibitemOpen
	\bibfield  {author} {\bibinfo {author} {\bibfnamefont {Athena~S.}\
			\bibnamefont {Sefat}},\ }\bibfield  {title} {\enquote {\bibinfo {title}
			{Pressure effects on two superconducting iron-based families},}\ }\href
	{\doibase 10.1088/0034-4885/74/12/124502} {\bibfield  {journal} {\bibinfo
			{journal} {Reports on Progress in Physics}\ }\textbf {\bibinfo {volume}
			{74}},\ \bibinfo {pages} {124502} (\bibinfo {year} {2011})}\BibitemShut
	{NoStop}%
	\bibitem [{\citenamefont {Iida}\ \emph {et~al.}(2009)\citenamefont {Iida},
		\citenamefont {H\"anisch}, \citenamefont {H\"uhne}, \citenamefont {Kurth},
		\citenamefont {Kidszun}, \citenamefont {Haindl}, \citenamefont {Werner},
		\citenamefont {Schultz},\ and\ \citenamefont {Holzapfel}}]{Iida2009}%
	\BibitemOpen
	\bibfield  {author} {\bibinfo {author} {\bibfnamefont {K.}~\bibnamefont
			{Iida}}, \bibinfo {author} {\bibfnamefont {J.}~\bibnamefont {H\"anisch}},
		\bibinfo {author} {\bibfnamefont {R.}~\bibnamefont {H\"uhne}}, \bibinfo
		{author} {\bibfnamefont {F.}~\bibnamefont {Kurth}}, \bibinfo {author}
		{\bibfnamefont {M.}~\bibnamefont {Kidszun}}, \bibinfo {author} {\bibfnamefont
			{S.}~\bibnamefont {Haindl}}, \bibinfo {author} {\bibfnamefont
			{J.}~\bibnamefont {Werner}}, \bibinfo {author} {\bibfnamefont
			{L.}~\bibnamefont {Schultz}}, \ and\ \bibinfo {author} {\bibfnamefont
			{B.}~\bibnamefont {Holzapfel}},\ }\bibfield  {title} {\enquote {\bibinfo
			{title} {Strong {$T_c$} dependence for strained epitaxial
				{Ba(Fe$_{1-x}$Co$_x$)$_2$As$_2$} thin films},}\ }\href {\doibase
		http://dx.doi.org/10.1063/1.3259922} {\bibfield  {journal} {\bibinfo
			{journal} {Applied Physics Letters}\ }\textbf {\bibinfo {volume} {95}},\
		\bibinfo {pages} {192501} (\bibinfo {year} {2009})}\BibitemShut {NoStop}%
	\bibitem [{\citenamefont {Engelmann}\ \emph {et~al.}(2013)\citenamefont
		{Engelmann}, \citenamefont {Grinenko}, \citenamefont {Chekhonin},
		\citenamefont {Skrotzki}, \citenamefont {Efremov}, \citenamefont {Oswald},
		\citenamefont {Iida}, \citenamefont {H\"uhne}, \citenamefont {H\"anisch},
		\citenamefont {Hoffmann}, \citenamefont {Kurth}, \citenamefont {Schultz},\
		and\ \citenamefont {Holzapfel}}]{Engelmann2013}%
	\BibitemOpen
	\bibfield  {author} {\bibinfo {author} {\bibfnamefont {J.}~\bibnamefont
			{Engelmann}}, \bibinfo {author} {\bibfnamefont {V.}~\bibnamefont {Grinenko}},
		\bibinfo {author} {\bibfnamefont {P.}~\bibnamefont {Chekhonin}}, \bibinfo
		{author} {\bibfnamefont {W.}~\bibnamefont {Skrotzki}}, \bibinfo {author}
		{\bibfnamefont {D.V.}\ \bibnamefont {Efremov}}, \bibinfo {author}
		{\bibfnamefont {S.}~\bibnamefont {Oswald}}, \bibinfo {author} {\bibfnamefont
			{K.}~\bibnamefont {Iida}}, \bibinfo {author} {\bibfnamefont {R.}~\bibnamefont
			{H\"uhne}}, \bibinfo {author} {\bibfnamefont {J.}~\bibnamefont {H\"anisch}},
		\bibinfo {author} {\bibfnamefont {M.}~\bibnamefont {Hoffmann}}, \bibinfo
		{author} {\bibfnamefont {F.}~\bibnamefont {Kurth}}, \bibinfo {author}
		{\bibfnamefont {L.}~\bibnamefont {Schultz}}, \ and\ \bibinfo {author}
		{\bibfnamefont {B.}~\bibnamefont {Holzapfel}},\ }\bibfield  {title} {\enquote
		{\bibinfo {title} {Strain induced superconductivity in the parent compound
				{BaFe$_2$As$_2$}},}\ }\href {\doibase 10.1038/ncomms3877} {\bibfield
		{journal} {\bibinfo  {journal} {Nat Commun}\ }\textbf {\bibinfo {volume} {4}}
		(\bibinfo {year} {2013}),\ 10.1038/ncomms3877}\BibitemShut {NoStop}%
	\bibitem [{\citenamefont {Iida}\ \emph {et~al.}(2016)\citenamefont {Iida},
		\citenamefont {Grinenko}, \citenamefont {Kurth}, \citenamefont {Ichinose},
		\citenamefont {Tsukada}, \citenamefont {Ahrens}, \citenamefont {Pukenas},
		\citenamefont {Chekhonin}, \citenamefont {Skrotzki}, \citenamefont
		{Teresiak}, \citenamefont {H\"uhne}, \citenamefont {Aswartham}, \citenamefont
		{Wurmehl}, \citenamefont {M\"onch}, \citenamefont {Erbe}, \citenamefont
		{H\"anisch}, \citenamefont {Holzapfel}, \citenamefont {Drechsler},\ and\
		\citenamefont {Efremov}}]{Kazumasa2016}%
	\BibitemOpen
	\bibfield  {author} {\bibinfo {author} {\bibfnamefont {Kazumasa}\
			\bibnamefont {Iida}}, \bibinfo {author} {\bibfnamefont {Vadim}\ \bibnamefont
			{Grinenko}}, \bibinfo {author} {\bibfnamefont {Fritz}\ \bibnamefont {Kurth}},
		\bibinfo {author} {\bibfnamefont {Ataru}\ \bibnamefont {Ichinose}}, \bibinfo
		{author} {\bibfnamefont {Ichiro}\ \bibnamefont {Tsukada}}, \bibinfo {author}
		{\bibfnamefont {Eike}\ \bibnamefont {Ahrens}}, \bibinfo {author}
		{\bibfnamefont {Aurimas}\ \bibnamefont {Pukenas}}, \bibinfo {author}
		{\bibfnamefont {Paul}\ \bibnamefont {Chekhonin}}, \bibinfo {author}
		{\bibfnamefont {Werner}\ \bibnamefont {Skrotzki}}, \bibinfo {author}
		{\bibfnamefont {Angelika}\ \bibnamefont {Teresiak}}, \bibinfo {author}
		{\bibfnamefont {Ruben}\ \bibnamefont {H\"uhne}}, \bibinfo {author}
		{\bibfnamefont {Saicharan}\ \bibnamefont {Aswartham}}, \bibinfo {author}
		{\bibfnamefont {Sabinea}\ \bibnamefont {Wurmehl}}, \bibinfo {author}
		{\bibfnamefont {Ingolf}\ \bibnamefont {M\"onch}}, \bibinfo {author}
		{\bibfnamefont {Manuela}\ \bibnamefont {Erbe}}, \bibinfo {author}
		{\bibfnamefont {Jens}\ \bibnamefont {H\"anisch}}, \bibinfo {author}
		{\bibfnamefont {Bernhard}\ \bibnamefont {Holzapfel}}, \bibinfo {author}
		{\bibfnamefont {Stefan-Ludwig}\ \bibnamefont {Drechsler}}, \ and\ \bibinfo
		{author} {\bibfnamefont {Dmitri~V.}\ \bibnamefont {Efremov}},\ }\bibfield
	{title} {\enquote {\bibinfo {title} {Hall-plot of the phase diagram for
				{Ba(Fe$_{1-x}$Co$_x$)$_2$As$_2$}},}\ }\href {\doibase 10.1038/srep28390}
	{\bibfield  {journal} {\bibinfo  {journal} {Scientific Reports}\ }\textbf
		{\bibinfo {volume} {6}} (\bibinfo {year} {2016}),\
		10.1038/srep28390}\BibitemShut {NoStop}%
	\bibitem [{\citenamefont {Torikachvili}\ \emph {et~al.}(2009)\citenamefont
		{Torikachvili}, \citenamefont {Bud'ko}, \citenamefont {Ni}, \citenamefont
		{Canfield},\ and\ \citenamefont {Hannahs}}]{Torikachvili2009}%
	\BibitemOpen
	\bibfield  {author} {\bibinfo {author} {\bibfnamefont {M.~S.}\ \bibnamefont
			{Torikachvili}}, \bibinfo {author} {\bibfnamefont {S.~L.}\ \bibnamefont
			{Bud'ko}}, \bibinfo {author} {\bibfnamefont {N.}~\bibnamefont {Ni}}, \bibinfo
		{author} {\bibfnamefont {P.~C.}\ \bibnamefont {Canfield}}, \ and\ \bibinfo
		{author} {\bibfnamefont {S.~T.}\ \bibnamefont {Hannahs}},\ }\bibfield
	{title} {\enquote {\bibinfo {title} {Effect of pressure on transport and
				magnetotransport properties in {CaFe$_{2}$As$_{2}$} single crystals},}\
	}\href {\doibase 10.1103/PhysRevB.80.014521} {\bibfield  {journal} {\bibinfo
			{journal} {Phys. Rev. B}\ }\textbf {\bibinfo {volume} {80}},\ \bibinfo
		{pages} {014521} (\bibinfo {year} {2009})}\BibitemShut {NoStop}%
	\bibitem [{\citenamefont {Proke\v{s}}\ \emph {et~al.}(2010)\citenamefont
		{Proke\v{s}}, \citenamefont {Kreyssig}, \citenamefont {Ouladdiaf},
		\citenamefont {Pratt}, \citenamefont {Ni}, \citenamefont {Bud'ko},
		\citenamefont {Canfield}, \citenamefont {McQueeney}, \citenamefont
		{Argyriou},\ and\ \citenamefont {Goldman}}]{Prokes2010}%
	\BibitemOpen
	\bibfield  {author} {\bibinfo {author} {\bibfnamefont {K.}~\bibnamefont
			{Proke\v{s}}}, \bibinfo {author} {\bibfnamefont {A.}~\bibnamefont
			{Kreyssig}}, \bibinfo {author} {\bibfnamefont {B.}~\bibnamefont {Ouladdiaf}},
		\bibinfo {author} {\bibfnamefont {D.~K.}\ \bibnamefont {Pratt}}, \bibinfo
		{author} {\bibfnamefont {N.}~\bibnamefont {Ni}}, \bibinfo {author}
		{\bibfnamefont {S.~L.}\ \bibnamefont {Bud'ko}}, \bibinfo {author}
		{\bibfnamefont {P.~C.}\ \bibnamefont {Canfield}}, \bibinfo {author}
		{\bibfnamefont {R.~J.}\ \bibnamefont {McQueeney}}, \bibinfo {author}
		{\bibfnamefont {D.~N.}\ \bibnamefont {Argyriou}}, \ and\ \bibinfo {author}
		{\bibfnamefont {A.~I.}\ \bibnamefont {Goldman}},\ }\bibfield  {title}
	{\enquote {\bibinfo {title} {Evidence from neutron diffraction for
				superconductivity in the stabilized tetragonal phase of {CaFe$_{2}$As$_{2}$}
				under uniaxial pressure},}\ }\href {\doibase 10.1103/PhysRevB.81.180506}
	{\bibfield  {journal} {\bibinfo  {journal} {Phys. Rev. B}\ }\textbf {\bibinfo
			{volume} {81}},\ \bibinfo {pages} {180506} (\bibinfo {year}
		{2010})}\BibitemShut {NoStop}%
	\bibitem [{\citenamefont {Bud'ko}\ \emph {et~al.}(2009)\citenamefont {Bud'ko},
		\citenamefont {Ni}, \citenamefont {Nandi}, \citenamefont {Schmiedeshoff},\
		and\ \citenamefont {Canfield}}]{Budko2009}%
	\BibitemOpen
	\bibfield  {author} {\bibinfo {author} {\bibfnamefont {S.~L.}\ \bibnamefont
			{Bud'ko}}, \bibinfo {author} {\bibfnamefont {N.}~\bibnamefont {Ni}}, \bibinfo
		{author} {\bibfnamefont {S.}~\bibnamefont {Nandi}}, \bibinfo {author}
		{\bibfnamefont {G.~M.}\ \bibnamefont {Schmiedeshoff}}, \ and\ \bibinfo
		{author} {\bibfnamefont {P.~C.}\ \bibnamefont {Canfield}},\ }\bibfield
	{title} {\enquote {\bibinfo {title} {Thermal expansion and anisotropic
				pressure derivatives of {$T_c$} in {Ba(Fe$_{1-x}$Co$_x$)$_2$As$_2$} single
				crystals},}\ }\href {\doibase 10.1103/PhysRevB.79.054525} {\bibfield
		{journal} {\bibinfo  {journal} {Phys. Rev. B}\ }\textbf {\bibinfo {volume}
			{79}},\ \bibinfo {pages} {054525} (\bibinfo {year} {2009})}\BibitemShut
	{NoStop}%
	\bibitem [{\citenamefont {Yamazaki}\ \emph {et~al.}(2010)\citenamefont
		{Yamazaki}, \citenamefont {Takeshita}, \citenamefont {Kobayashi},
		\citenamefont {Fukazawa}, \citenamefont {Kohori}, \citenamefont {Kihou},
		\citenamefont {Lee}, \citenamefont {Kito}, \citenamefont {Iyo},\ and\
		\citenamefont {Eisaki}}]{Yamazaki2010}%
	\BibitemOpen
	\bibfield  {author} {\bibinfo {author} {\bibfnamefont {T.}~\bibnamefont
			{Yamazaki}}, \bibinfo {author} {\bibfnamefont {N.}~\bibnamefont {Takeshita}},
		\bibinfo {author} {\bibfnamefont {R.}~\bibnamefont {Kobayashi}}, \bibinfo
		{author} {\bibfnamefont {H.}~\bibnamefont {Fukazawa}}, \bibinfo {author}
		{\bibfnamefont {Y.}~\bibnamefont {Kohori}}, \bibinfo {author} {\bibfnamefont
			{K.}~\bibnamefont {Kihou}}, \bibinfo {author} {\bibfnamefont {C.-H.}\
			\bibnamefont {Lee}}, \bibinfo {author} {\bibfnamefont {H.}~\bibnamefont
			{Kito}}, \bibinfo {author} {\bibfnamefont {A.}~\bibnamefont {Iyo}}, \ and\
		\bibinfo {author} {\bibfnamefont {H.}~\bibnamefont {Eisaki}},\ }\bibfield
	{title} {\enquote {\bibinfo {title} {Appearance of pressure-induced
				superconductivity in {B}a{Fe}$_{2}${A}s$_{2}$ under
				hydrostatic conditions and its extremely high sensitivity to uniaxial
				stress},}\ }\href@noop {} {\bibfield  {journal} {\bibinfo  {journal} {Phys.
				Rev. B}\ }\textbf {\bibinfo {volume} {81}},\ \bibinfo {pages} {224511}
		(\bibinfo {year} {2010})}\BibitemShut {NoStop}%
	\bibitem [{\citenamefont {Meingast}\ \emph {et~al.}(2012)\citenamefont
		{Meingast}, \citenamefont {Hardy}, \citenamefont {Heid}, \citenamefont
		{Adelmann}, \citenamefont {B\"ohmer}, \citenamefont {Burger}, \citenamefont
		{Ernst}, \citenamefont {Fromknecht}, \citenamefont {Schweiss},\ and\
		\citenamefont {Wolf}}]{Meingast2012}%
	\BibitemOpen
	\bibfield  {author} {\bibinfo {author} {\bibfnamefont {Christoph}\
			\bibnamefont {Meingast}}, \bibinfo {author} {\bibfnamefont {Fr\'ed\'eric}\
			\bibnamefont {Hardy}}, \bibinfo {author} {\bibfnamefont {Rolf}\ \bibnamefont
			{Heid}}, \bibinfo {author} {\bibfnamefont {Peter}\ \bibnamefont {Adelmann}},
		\bibinfo {author} {\bibfnamefont {Anna}\ \bibnamefont {B\"ohmer}}, \bibinfo
		{author} {\bibfnamefont {Philipp}\ \bibnamefont {Burger}}, \bibinfo {author}
		{\bibfnamefont {Doris}\ \bibnamefont {Ernst}}, \bibinfo {author}
		{\bibfnamefont {Rainer}\ \bibnamefont {Fromknecht}}, \bibinfo {author}
		{\bibfnamefont {Peter}\ \bibnamefont {Schweiss}}, \ and\ \bibinfo {author}
		{\bibfnamefont {Thomas}\ \bibnamefont {Wolf}},\ }\bibfield  {title} {\enquote
		{\bibinfo {title} {Thermal expansion and {G}r\"{u}neisen parameters of
				{Ba(Fe$_{1-x}$Co$_x$)$_2$As$_2$} - a thermodynamic quest for quantum
				criticality},}\ }\href {\doibase 10.1103/PhysRevLett.108.177004} {\bibfield
		{journal} {\bibinfo  {journal} {Phys. Rev. Lett.}\ }\textbf {\bibinfo
			{volume} {108}},\ \bibinfo {pages} {177004} (\bibinfo {year}
		{2012})}\BibitemShut {NoStop}%
	\bibitem [{\citenamefont {Ran}\ \emph {et~al.}(2012)\citenamefont {Ran},
		\citenamefont {Bud'ko}, \citenamefont {Straszheim}, \citenamefont {Soh},
		\citenamefont {Kim}, \citenamefont {Kreyssig}, \citenamefont {Goldman},\ and\
		\citenamefont {Canfield}}]{Ran2012}%
	\BibitemOpen
	\bibfield  {author} {\bibinfo {author} {\bibfnamefont {S.}~\bibnamefont
			{Ran}}, \bibinfo {author} {\bibfnamefont {S.~L.}\ \bibnamefont {Bud'ko}},
		\bibinfo {author} {\bibfnamefont {W.~E.}\ \bibnamefont {Straszheim}},
		\bibinfo {author} {\bibfnamefont {J.}~\bibnamefont {Soh}}, \bibinfo {author}
		{\bibfnamefont {M.~G.}\ \bibnamefont {Kim}}, \bibinfo {author} {\bibfnamefont
			{A.}~\bibnamefont {Kreyssig}}, \bibinfo {author} {\bibfnamefont {A.~I.}\
			\bibnamefont {Goldman}}, \ and\ \bibinfo {author} {\bibfnamefont {P.~C.}\
			\bibnamefont {Canfield}},\ }\bibfield  {title} {\enquote {\bibinfo {title}
			{Control of magnetic, nonmagnetic, and superconducting states in annealed
				{Ca(Fe$_{1-x}$Co$_{x}$)$_{2}$As$_{2}$}},}\ }\href {\doibase
		10.1103/PhysRevB.85.224528} {\bibfield  {journal} {\bibinfo  {journal} {Phys.
				Rev. B}\ }\textbf {\bibinfo {volume} {85}},\ \bibinfo {pages} {224528}
		(\bibinfo {year} {2012})}\BibitemShut {NoStop}%
	\bibitem [{\citenamefont {Gati}\ \emph {et~al.}(2012)\citenamefont {Gati},
		\citenamefont {K{\"o}hler}, \citenamefont {Guterding}, \citenamefont {Wolf},
		\citenamefont {Kn{\"o}ner}, \citenamefont {Ran}, \citenamefont {Bud{'}ko},
		\citenamefont {Canfield},\ and\ \citenamefont {Lang}}]{Gati2012}%
	\BibitemOpen
	\bibfield  {author} {\bibinfo {author} {\bibfnamefont {E.}~\bibnamefont
			{Gati}}, \bibinfo {author} {\bibfnamefont {S.}~\bibnamefont {K{\"o}hler}},
		\bibinfo {author} {\bibfnamefont {D.}~\bibnamefont {Guterding}}, \bibinfo
		{author} {\bibfnamefont {B.}~\bibnamefont {Wolf}}, \bibinfo {author}
		{\bibfnamefont {S.}~\bibnamefont {Kn{\"o}ner}}, \bibinfo {author}
		{\bibfnamefont {S.}~\bibnamefont {Ran}}, \bibinfo {author} {\bibfnamefont
			{S.~L.}\ \bibnamefont {Bud{'}ko}}, \bibinfo {author} {\bibfnamefont {P.~C.}\
			\bibnamefont {Canfield}}, \ and\ \bibinfo {author} {\bibfnamefont
			{M.}~\bibnamefont {Lang}},\ }\bibfield  {title} {\enquote {\bibinfo {title}
			{Hydrostatic-pressure tuning of magnetic, nonmagnetic, and superconducting
				states in annealed {Ca(Fe$_{1-x}$Co$_x$)$_{2}$As$_{2}$}},}\ }\href {\doibase
		10.1103/PhysRevB.86.220511} {\bibfield  {journal} {\bibinfo  {journal} {Phys.
				Rev. B}\ }\textbf {\bibinfo {volume} {86}},\ \bibinfo {pages} {220511}
		(\bibinfo {year} {2012})}\BibitemShut {NoStop}%
	\bibitem [{\citenamefont {Bud'ko}\ \emph {et~al.}(2013)\citenamefont {Bud'ko},
		\citenamefont {Ran},\ and\ \citenamefont {Canfield}}]{Budko2012}%
	\BibitemOpen
	\bibfield  {author} {\bibinfo {author} {\bibfnamefont {Sergey~L.}\
			\bibnamefont {Bud'ko}}, \bibinfo {author} {\bibfnamefont {Sheng}\
			\bibnamefont {Ran}}, \ and\ \bibinfo {author} {\bibfnamefont {Paul~C.}\
			\bibnamefont {Canfield}},\ }\bibfield  {title} {\enquote {\bibinfo {title}
			{Thermal expansion of {CaFe$_{2}$As$_{2}$}: Effect of cobalt doping and
				postgrowth thermal treatment},}\ }\href {\doibase 10.1103/PhysRevB.88.064513}
	{\bibfield  {journal} {\bibinfo  {journal} {Phys. Rev. B}\ }\textbf {\bibinfo
			{volume} {88}},\ \bibinfo {pages} {064513} (\bibinfo {year}
		{2013})}\BibitemShut {NoStop}%
	\bibitem [{\citenamefont {Colombier}\ \emph {et~al.}(2009)\citenamefont
		{Colombier}, \citenamefont {Bud'ko}, \citenamefont {Ni},\ and\ \citenamefont
		{Canfield}}]{Colombier2009}%
	\BibitemOpen
	\bibfield  {author} {\bibinfo {author} {\bibfnamefont {E.}~\bibnamefont
			{Colombier}}, \bibinfo {author} {\bibfnamefont {S.~L.}\ \bibnamefont
			{Bud'ko}}, \bibinfo {author} {\bibfnamefont {N.}~\bibnamefont {Ni}}, \ and\
		\bibinfo {author} {\bibfnamefont {P.~C.}\ \bibnamefont {Canfield}},\
	}\bibfield  {title} {\enquote {\bibinfo {title} {Complete pressure-dependent
				phase diagrams for {SrFe$_{2}$As$_{2}$} and {BaFe$_{2}$As$_{2}$}},}\ }\href
	{\doibase 10.1103/PhysRevB.79.224518} {\bibfield  {journal} {\bibinfo
			{journal} {Phys. Rev. B}\ }\textbf {\bibinfo {volume} {79}},\ \bibinfo
		{pages} {224518} (\bibinfo {year} {2009})}\BibitemShut {NoStop}%
	\bibitem [{\citenamefont {Ran}\ \emph {et~al.}(2011)\citenamefont {Ran},
		\citenamefont {Bud'ko}, \citenamefont {Pratt}, \citenamefont {Kreyssig},
		\citenamefont {Kim}, \citenamefont {Kramer}, \citenamefont {Ryan},
		\citenamefont {Rowan-Weetaluktuk}, \citenamefont {Furukawa}, \citenamefont
		{Roy}, \citenamefont {Goldman},\ and\ \citenamefont {Canfield}}]{Ran2011}%
	\BibitemOpen
	\bibfield  {author} {\bibinfo {author} {\bibfnamefont {S.}~\bibnamefont
			{Ran}}, \bibinfo {author} {\bibfnamefont {S.~L.}\ \bibnamefont {Bud'ko}},
		\bibinfo {author} {\bibfnamefont {D.~K.}\ \bibnamefont {Pratt}}, \bibinfo
		{author} {\bibfnamefont {A.}~\bibnamefont {Kreyssig}}, \bibinfo {author}
		{\bibfnamefont {M.~G.}\ \bibnamefont {Kim}}, \bibinfo {author} {\bibfnamefont
			{M.~J.}\ \bibnamefont {Kramer}}, \bibinfo {author} {\bibfnamefont {D.~H.}\
			\bibnamefont {Ryan}}, \bibinfo {author} {\bibfnamefont {W.~N.}\ \bibnamefont
			{Rowan-Weetaluktuk}}, \bibinfo {author} {\bibfnamefont {Y.}~\bibnamefont
			{Furukawa}}, \bibinfo {author} {\bibfnamefont {B.}~\bibnamefont {Roy}},
		\bibinfo {author} {\bibfnamefont {A.~I.}\ \bibnamefont {Goldman}}, \ and\
		\bibinfo {author} {\bibfnamefont {P.~C.}\ \bibnamefont {Canfield}},\
	}\bibfield  {title} {\enquote {\bibinfo {title} {Stabilization of an
				ambient-pressure collapsed tetragonal phase in {CaFe$_{2}$As$_{2}$} and
				tuning of the orthorhombic-antiferromagnetic transition temperature by over
				{70 K} via control of nanoscale precipitates},}\ }\href {\doibase
		10.1103/PhysRevB.83.144517} {\bibfield  {journal} {\bibinfo  {journal} {Phys.
				Rev. B}\ }\textbf {\bibinfo {volume} {83}},\ \bibinfo {pages} {144517}
		(\bibinfo {year} {2011})}\BibitemShut {NoStop}%
	\bibitem [{\citenamefont {Ran}(2014)}]{Ranthesis}%
	\BibitemOpen
	\bibfield  {author} {\bibinfo {author} {\bibfnamefont {Sheng}\ \bibnamefont
			{Ran}},\ }\emph {\bibinfo {title} {Combined effects of post-growth thermal
			treatment and chemical substitution on physical properties of
			{CaFe$_2$As$_2$}}},\ \href@noop {} {Ph.D. thesis},\ \bibinfo  {school} {Iowa
		State University} (\bibinfo {year} {2014})\BibitemShut {NoStop}%
	\bibitem [{\citenamefont {Fisher}\ \emph {et~al.}(2011)\citenamefont {Fisher},
		\citenamefont {Degiorgi},\ and\ \citenamefont {Shen}}]{Fisher2011}%
	\BibitemOpen
	\bibfield  {author} {\bibinfo {author} {\bibfnamefont {I~R}\ \bibnamefont
			{Fisher}}, \bibinfo {author} {\bibfnamefont {L}~\bibnamefont {Degiorgi}}, \
		and\ \bibinfo {author} {\bibfnamefont {Z~X}\ \bibnamefont {Shen}},\
	}\bibfield  {title} {\enquote {\bibinfo {title} {In-plane electronic
				anisotropy of underdoped '122' {Fe}-arsenide superconductors revealed by
				measurements of detwinned single crystals},}\ }\href
	{http://stacks.iop.org/0034-4885/74/i=12/a=124506} {\bibfield  {journal}
		{\bibinfo  {journal} {Reports on Progress in Physics}\ }\textbf {\bibinfo
			{volume} {74}},\ \bibinfo {pages} {124506} (\bibinfo {year}
		{2011})}\BibitemShut {NoStop}%
	\bibitem [{\citenamefont {Sapkota}\ \emph {et~al.}(2014)\citenamefont
		{Sapkota}, \citenamefont {Tucker}, \citenamefont {Ramazanoglu}, \citenamefont
		{Tian}, \citenamefont {Ni}, \citenamefont {Cava}, \citenamefont {McQueeney},
		\citenamefont {Goldman},\ and\ \citenamefont {Kreyssig}}]{Sapkota2014}%
	\BibitemOpen
	\bibfield  {author} {\bibinfo {author} {\bibfnamefont {A.}~\bibnamefont
			{Sapkota}}, \bibinfo {author} {\bibfnamefont {G.~S.}\ \bibnamefont {Tucker}},
		\bibinfo {author} {\bibfnamefont {M.}~\bibnamefont {Ramazanoglu}}, \bibinfo
		{author} {\bibfnamefont {W.}~\bibnamefont {Tian}}, \bibinfo {author}
		{\bibfnamefont {N.}~\bibnamefont {Ni}}, \bibinfo {author} {\bibfnamefont
			{R.~J.}\ \bibnamefont {Cava}}, \bibinfo {author} {\bibfnamefont {R.~J.}\
			\bibnamefont {McQueeney}}, \bibinfo {author} {\bibfnamefont {A.~I.}\
			\bibnamefont {Goldman}}, \ and\ \bibinfo {author} {\bibfnamefont
			{A.}~\bibnamefont {Kreyssig}},\ }\bibfield  {title} {\enquote {\bibinfo
			{title} {Lattice distortion and stripelike antiferromagnetic order in
				{Ca$_{10}$(Pt$_{3}$As$_{8}$)(Fe$_{2}$As$_{2}$)$_{5}$}},}\ }\href {\doibase
		10.1103/PhysRevB.90.100504} {\bibfield  {journal} {\bibinfo  {journal} {Phys.
				Rev. B}\ }\textbf {\bibinfo {volume} {90}},\ \bibinfo {pages} {100504}
		(\bibinfo {year} {2014})}\BibitemShut {NoStop}%
	\bibitem [{\citenamefont {Schmiedeshoff}\ \emph {et~al.}(2006)\citenamefont
		{Schmiedeshoff}, \citenamefont {Lounsbury}, \citenamefont {Luna},
		\citenamefont {Tracy}, \citenamefont {Schramm}, \citenamefont {Tozer},
		\citenamefont {Correa}, \citenamefont {Hannahs}, \citenamefont {Murphy},
		\citenamefont {Palm}, \citenamefont {Lacerda}, \citenamefont {Bud’ko},
		\citenamefont {Canfield}, \citenamefont {Smith}, \citenamefont {Lashley},\
		and\ \citenamefont {Cooley}}]{Schmiedeshoff2006}%
	\BibitemOpen
	\bibfield  {author} {\bibinfo {author} {\bibfnamefont {G.~M.}\ \bibnamefont
			{Schmiedeshoff}}, \bibinfo {author} {\bibfnamefont {A.~W.}\ \bibnamefont
			{Lounsbury}}, \bibinfo {author} {\bibfnamefont {D.~J.}\ \bibnamefont {Luna}},
		\bibinfo {author} {\bibfnamefont {S.~J.}\ \bibnamefont {Tracy}}, \bibinfo
		{author} {\bibfnamefont {A.~J.}\ \bibnamefont {Schramm}}, \bibinfo {author}
		{\bibfnamefont {S.~W.}\ \bibnamefont {Tozer}}, \bibinfo {author}
		{\bibfnamefont {V.~F.}\ \bibnamefont {Correa}}, \bibinfo {author}
		{\bibfnamefont {S.~T.}\ \bibnamefont {Hannahs}}, \bibinfo {author}
		{\bibfnamefont {T.~P.}\ \bibnamefont {Murphy}}, \bibinfo {author}
		{\bibfnamefont {E.~C.}\ \bibnamefont {Palm}}, \bibinfo {author}
		{\bibfnamefont {A.~H.}\ \bibnamefont {Lacerda}}, \bibinfo {author}
		{\bibfnamefont {S.~L.}\ \bibnamefont {Bud’ko}}, \bibinfo {author}
		{\bibfnamefont {P.~C.}\ \bibnamefont {Canfield}}, \bibinfo {author}
		{\bibfnamefont {J.~L.}\ \bibnamefont {Smith}}, \bibinfo {author}
		{\bibfnamefont {J.~C.}\ \bibnamefont {Lashley}}, \ and\ \bibinfo {author}
		{\bibfnamefont {J.~C.}\ \bibnamefont {Cooley}},\ }\bibfield  {title}
	{\enquote {\bibinfo {title} {Versatile and compact capacitive dilatometer},}\
	}\href {\doibase http://dx.doi.org/10.1063/1.2403088} {\bibfield  {journal}
		{\bibinfo  {journal} {Review of Scientific Instruments}\ }\textbf {\bibinfo
			{volume} {77}},\ \bibinfo {pages} {123907} (\bibinfo {year}
		{2006})}\BibitemShut {NoStop}%
	\bibitem [{\citenamefont {Tomi{\'{c}}}\ \emph {et~al.}(2012)\citenamefont
		{Tomi{\'{c}}}, \citenamefont {Valent{\'{i}}},\ and\ \citenamefont
		{Jeschke}}]{Tomic2012}%
	\BibitemOpen
	\bibfield  {author} {\bibinfo {author} {\bibfnamefont {M.}~\bibnamefont
			{Tomi{\'{c}}}}, \bibinfo {author} {\bibfnamefont {R.}~\bibnamefont
			{Valent{\'{i}}}}, \ and\ \bibinfo {author} {\bibfnamefont {H.~O.}\
			\bibnamefont {Jeschke}},\ }\bibfield  {title} {\enquote {\bibinfo {title}
			{Uniaxial versus hydrostatic pressure-induced phase transitions in
				{CaFe$_2$As$_2$} and {BaFe$_2$As$_2$}},}\ }\href@noop {} {\bibfield
		{journal} {\bibinfo  {journal} {Phys. Rev. B}\ }\textbf {\bibinfo {volume}
			{85}},\ \bibinfo {pages} {094105} (\bibinfo {year} {2012})}\BibitemShut
	{NoStop}%
	\bibitem [{\citenamefont {Tanatar}\ \emph {et~al.}(2009)\citenamefont
		{Tanatar}, \citenamefont {Kreyssig}, \citenamefont {Nandi}, \citenamefont
		{Ni}, \citenamefont {Bud'ko}, \citenamefont {Canfield}, \citenamefont
		{Goldman},\ and\ \citenamefont {Prozorov}}]{Tanatar2009}%
	\BibitemOpen
	\bibfield  {author} {\bibinfo {author} {\bibfnamefont {M.~A.}\ \bibnamefont
			{Tanatar}}, \bibinfo {author} {\bibfnamefont {A.}~\bibnamefont {Kreyssig}},
		\bibinfo {author} {\bibfnamefont {S.}~\bibnamefont {Nandi}}, \bibinfo
		{author} {\bibfnamefont {N.}~\bibnamefont {Ni}}, \bibinfo {author}
		{\bibfnamefont {S.~L.}\ \bibnamefont {Bud'ko}}, \bibinfo {author}
		{\bibfnamefont {P.~C.}\ \bibnamefont {Canfield}}, \bibinfo {author}
		{\bibfnamefont {A.~I.}\ \bibnamefont {Goldman}}, \ and\ \bibinfo {author}
		{\bibfnamefont {R.}~\bibnamefont {Prozorov}},\ }\bibfield  {title} {\enquote
		{\bibinfo {title} {Direct imaging of the structural domains in the iron
				pnictides {$A$Fe$_2$As$_2$ ($A$=Ca,Sr,Ba)}},}\ }\href {\doibase
		10.1103/PhysRevB.79.180508} {\bibfield  {journal} {\bibinfo  {journal} {Phys.
				Rev. B}\ }\textbf {\bibinfo {volume} {79}},\ \bibinfo {pages} {180508}
		(\bibinfo {year} {2009})}\BibitemShut {NoStop}%
	\bibitem [{\citenamefont {Goldman}\ \emph {et~al.}(2008)\citenamefont
		{Goldman}, \citenamefont {Argyriou}, \citenamefont {Ouladdiaf}, \citenamefont
		{Chatterji}, \citenamefont {Kreyssig}, \citenamefont {Nandi}, \citenamefont
		{Ni}, \citenamefont {Bud'ko}, \citenamefont {Canfield},\ and\ \citenamefont
		{McQueeney}}]{Goldman2008}%
	\BibitemOpen
	\bibfield  {author} {\bibinfo {author} {\bibfnamefont {A.~I.}\ \bibnamefont
			{Goldman}}, \bibinfo {author} {\bibfnamefont {D.~N.}\ \bibnamefont
			{Argyriou}}, \bibinfo {author} {\bibfnamefont {B.}~\bibnamefont {Ouladdiaf}},
		\bibinfo {author} {\bibfnamefont {T.}~\bibnamefont {Chatterji}}, \bibinfo
		{author} {\bibfnamefont {A.}~\bibnamefont {Kreyssig}}, \bibinfo {author}
		{\bibfnamefont {S.}~\bibnamefont {Nandi}}, \bibinfo {author} {\bibfnamefont
			{N.}~\bibnamefont {Ni}}, \bibinfo {author} {\bibfnamefont {S.~L.}\
			\bibnamefont {Bud'ko}}, \bibinfo {author} {\bibfnamefont {P.~C.}\
			\bibnamefont {Canfield}}, \ and\ \bibinfo {author} {\bibfnamefont {R.~J.}\
			\bibnamefont {McQueeney}},\ }\bibfield  {title} {\enquote {\bibinfo {title}
			{Lattice and magnetic instabilities in {CaFe$_{2}$As$_{2}$}: {A}
				single-crystal neutron diffraction study},}\ }\href {\doibase
		10.1103/PhysRevB.78.100506} {\bibfield  {journal} {\bibinfo  {journal} {Phys.
				Rev. B}\ }\textbf {\bibinfo {volume} {78}},\ \bibinfo {pages} {100506}
		(\bibinfo {year} {2008})}\BibitemShut {NoStop}%
	\bibitem [{\citenamefont {B{\"o}hmer}\ \emph {et~al.}(2015)\citenamefont
		{B{\"o}hmer}, \citenamefont {Hardy}, \citenamefont {Wang}, \citenamefont
		{Wolf}, \citenamefont {Schweiss},\ and\ \citenamefont
		{Meingast}}]{Boehmer2015II}%
	\BibitemOpen
	\bibfield  {author} {\bibinfo {author} {\bibfnamefont {A.~E.}\ \bibnamefont
			{B{\"o}hmer}}, \bibinfo {author} {\bibfnamefont {F.}~\bibnamefont {Hardy}},
		\bibinfo {author} {\bibfnamefont {L.}~\bibnamefont {Wang}}, \bibinfo {author}
		{\bibfnamefont {T.}~\bibnamefont {Wolf}}, \bibinfo {author} {\bibfnamefont
			{P.}~\bibnamefont {Schweiss}}, \ and\ \bibinfo {author} {\bibfnamefont
			{C.}~\bibnamefont {Meingast}},\ }\bibfield  {title} {\enquote {\bibinfo
			{title} {Superconductivity-induced reentrance of orthorhombic distortion in
				{Ba$_{1-x}$K$_x$Fe$_2$As$_2$}},}\ }\href {\doibase 10.1038/ncomms8911}
	{\bibfield  {journal} {\bibinfo  {journal} {Nature Communications}\ }\textbf
		{\bibinfo {volume} {6}},\ \bibinfo {pages} {8911} (\bibinfo {year}
		{2015})}\BibitemShut {NoStop}%
	\bibitem [{\citenamefont {Ni}\ \emph {et~al.}(2009)\citenamefont {Ni},
		\citenamefont {Thaler}, \citenamefont {Kracher}, \citenamefont {Yan},
		\citenamefont {Bud'ko},\ and\ \citenamefont {Canfield}}]{Ni2009}%
	\BibitemOpen
	\bibfield  {author} {\bibinfo {author} {\bibfnamefont {N.}~\bibnamefont
			{Ni}}, \bibinfo {author} {\bibfnamefont {A.}~\bibnamefont {Thaler}}, \bibinfo
		{author} {\bibfnamefont {A.}~\bibnamefont {Kracher}}, \bibinfo {author}
		{\bibfnamefont {J.~Q.}\ \bibnamefont {Yan}}, \bibinfo {author} {\bibfnamefont
			{S.~L.}\ \bibnamefont {Bud'ko}}, \ and\ \bibinfo {author} {\bibfnamefont
			{P.~C.}\ \bibnamefont {Canfield}},\ }\bibfield  {title} {\enquote {\bibinfo
			{title} {Phase diagrams of {Ba(Fe$_{1-x}$$M$$_x$)$_2$As$_{2}$} single
				crystals {($M$=Rh and Pd)}},}\ }\href {\doibase 10.1103/PhysRevB.80.024511}
	{\bibfield  {journal} {\bibinfo  {journal} {Phys. Rev. B}\ }\textbf {\bibinfo
			{volume} {80}},\ \bibinfo {pages} {024511} (\bibinfo {year}
		{2009})}\BibitemShut {NoStop}%
	\bibitem [{\citenamefont {Thaler}\ \emph {et~al.}(2010)\citenamefont {Thaler},
		\citenamefont {Ni}, \citenamefont {Kracher}, \citenamefont {Yan},
		\citenamefont {Bud'ko},\ and\ \citenamefont {Canfield}}]{Thaler2010}%
	\BibitemOpen
	\bibfield  {author} {\bibinfo {author} {\bibfnamefont {A.}~\bibnamefont
			{Thaler}}, \bibinfo {author} {\bibfnamefont {N.}~\bibnamefont {Ni}}, \bibinfo
		{author} {\bibfnamefont {A.}~\bibnamefont {Kracher}}, \bibinfo {author}
		{\bibfnamefont {J.~Q.}\ \bibnamefont {Yan}}, \bibinfo {author} {\bibfnamefont
			{S.~L.}\ \bibnamefont {Bud'ko}}, \ and\ \bibinfo {author} {\bibfnamefont
			{P.~C.}\ \bibnamefont {Canfield}},\ }\bibfield  {title} {\enquote {\bibinfo
			{title} {Physical and magnetic properties of
				{Ba(Fe$_{1-x}$Ru$_{x}$)$_{2}$As$_{2}$} single crystals},}\ }\href {\doibase
		10.1103/PhysRevB.82.014534} {\bibfield  {journal} {\bibinfo  {journal} {Phys.
				Rev. B}\ }\textbf {\bibinfo {volume} {82}},\ \bibinfo {pages} {014534}
		(\bibinfo {year} {2010})}\BibitemShut {NoStop}%
\end{thebibliography}

\end{document}